\documentclass[
  aps,
  prx,
  reprint,
  superscriptaddress,
  longbibliography,
  floatfix
]{revtex4-2}

\usepackage{amsmath,amssymb,amsfonts}
\usepackage{graphicx}
\usepackage{textcomp}
\usepackage{xcolor}
\usepackage{siunitx}
\usepackage[dvipsnames]{xcolor}
\usepackage{hyperref}
\usepackage[percent]{overpic}
\definecolor{QNT_green}{HTML}{539E8E}

\hypersetup{
  colorlinks   = true, %Colours links instead of ugly boxes
  urlcolor     = QNT_green, %Colour for external hyperlinks
  linkcolor    = QNT_green, %Colour of internal links
  citecolor   = QNT_green %Colour of citations
}

\begin{document}

\title{Generative IQP Circuit Learning with Physics-Informed Latent Initialization}

\author{Chen-Yu Liu}
\affiliation{Quantinuum, Partnership House, London, UK}

\author{Leonardo Placidi}
\affiliation{Quantinuum, Otemachi Financial City Grand Cube, Tokyo, Japan}

\author{Marco Ballarin}
\affiliation{Quantinuum, Partnership House, London, UK}

\author{Enrico Rinaldi}
\affiliation{Quantinuum, Partnership House, London, UK}

\begin{abstract}
Quantum generative learning based on instantaneous quantum polynomial-time (IQP) circuits can benefit from efficient classical training strategies.
A recent latent adaptation framework for IQP-based generative modeling transfers shared circuit parameters across instances of the same task with different hyperparameters while adapting only a low-dimensional latent variable for each new instance. However, existing approaches initialize this latent variable randomly, which can limit optimization efficiency and performance. In this work, we introduce a physics-informed latent initialization scheme for IQP generative learning to improve upon existing random initialization schemes. Motivated by the platonic representation hypothesis, we use latent representations extracted from a classical physics-informed neural network (PINN) surrogate to initialize the latent variables of the quantum model for the solution of the Burgers' equation. The initialized IQP model is then adapted on a higher-resolution solution domain. We find that this structured initialization consistently outperforms random latent initialization, yielding improved adaptation behavior and stronger generative accuracy across multiple viscosity settings. These results show that classical surrogate representations can provide useful inductive bias for quantum generative models and offer a practical route to improved initialization in IQP-based learning.
\end{abstract}

\maketitle

\section{Introduction}

Quantum circuit families whose output distributions are believed to be classically hard to sample from provide a natural setting in which to investigate practical routes toward quantum utility and quantum advantage, particularly in quantum machine learning \cite{schuld2019quantum, benedetti2019parameterized, mari2020transfer, liu2025quantum, huang2021power, belis2026spectral, 
pmlr-v322-liu26b,dong2008quantum,
liu2025measureoncedesigningsingleshot,
chen2020variational, liu2024training, lin2024quantum, hanruiwang2022quantumnas, du2021grover, liu2025a} and quantum generative modeling \cite{zhang2024generative, lloyd2018quantum, huang2025generative, recio2025train} . A prominent example is given by instantaneous quantum polynomial-time (IQP) circuits \cite{placidi2026impact, ballo2026shallow, liu2026toward, bremner2011classical, leontica2024exploring}, which form a restricted but nontrivial class of commuting quantum circuits whose output distributions are widely studied in the context of sampling hardness.
At the same time, the practical usefulness of IQP-based models depends not only on their expressive or complexity-theoretic properties, but also on whether they can be trained efficiently and robustly in settings relevant to scientific data. This challenge is particularly critical when the target data arise from continuous physical systems \cite{coyle2020born, shen2026characterizing, khojasteh2022lagrangian}, since one should simultaneously address data representation and scalable training objectives challenges.

A recent latent-adaptation framework for IQP generative modeling provides one such route \cite{liu2026toward}. In that approach, a shared circuit core is learned for a reference instance, while a low-dimensional latent parameter is adapted across different parameter regimes of the same task. This strategy is appealing both computationally and physically: it reduces the number of degrees of freedom that must be reoptimized for each new instance, and it organizes the learned family of solutions as a trajectory in a low-dimensional latent space. It also offers a pragmatic mechanism for extending quantum generative modeling to parameterized scientific systems, where neighboring instances often exhibit structured continuity. However, existing latent-adaptation schemes initialize the latent variable randomly. In a framework where only a small subset of parameters remains trainable, such initialization can affect optimization stability, convergence behavior, and final generation quality.  

The importance of initialization is well known more broadly in variational quantum algorithms \cite{mcclean2018barren, grant2019initialization,PuEtAl25,avqe}. The quantum approximate optimization algorithm (QAOA), for example, can be warm-started using classical preprocessing, such as semidefinite-programming relaxations \cite{egger2021warm}, to bias or initialize the quantum state and variational parameters, often yielding improved low-depth performance relative to standard QAOA when solving optimization problems. This illustrates a broader principle: classical information can be used to place a variational quantum model in a more favorable region of parameter space before quantum optimization begins. In parallel, physics-informed machine learning has developed a complementary set of ideas on the classical side. Physics-informed neural networks (PINNs) and related models incorporate governing differential equations directly into the learning objective \cite{raissi2019physics, karniadakis2021physics, park2024parameterized}, enabling compact latent representations for families of partial differential equation (PDE) solutions \cite{zhou2024multi}.

These observations motivate the central question of this work: can a physics-informed classical surrogate provide a useful initialization for IQP latent adaptation? We investigate this question in the setting of parameterized PDE solution families, using Burgers' equation \cite{raissi2019physics} as a representative and elementary benchmark. Our approach is to train a latent-conditioned PINN surrogate on lower-resolution solutions, extract the corresponding latent representation for a given viscosity instance, and transfer that latent as the initialization of the IQP latent variable for the corresponding higher-resolution generative task. The subsequent IQP optimization then proceeds within the standard latent-adaptation framework, with the transferred latent supplying a structured starting point in place of random initialization.

\begin{figure*}[t]
    \includegraphics[width=1\textwidth]{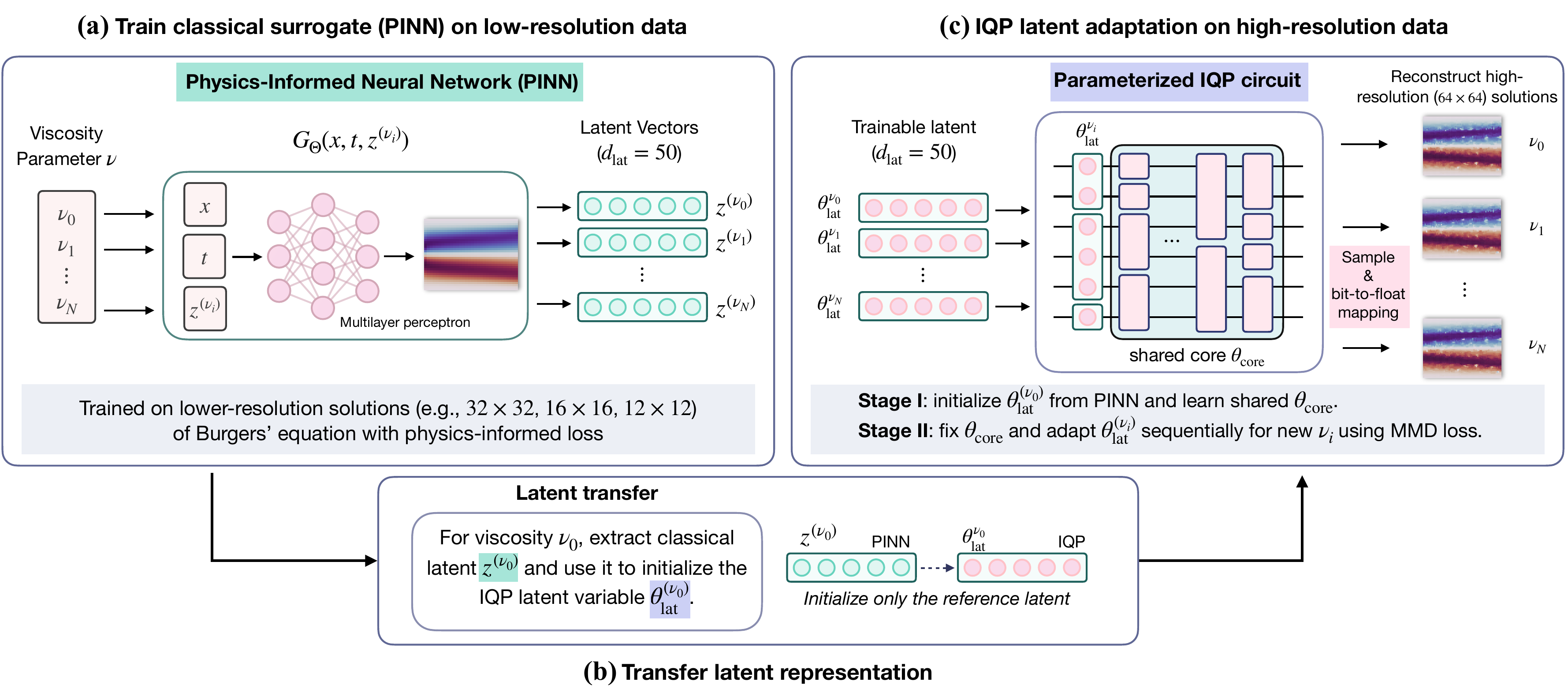}
    \caption{
    Schematic overview of the proposed physics-informed latent initialization scheme for IQP generative learning. 
    (a) A latent-conditioned PINN surrogate is trained on lower-resolution Burgers’ equation solution data, using one latent vector \(z^{(\nu_i)}\) for each viscosity instance \(\nu_i\) (Sec.~\ref{sec:pinn}). 
    (b) For the reference viscosity \(\nu_0\), the corresponding classical latent representation \(z^{(\nu_0)}\) is transferred to initialize the IQP latent variable \(\theta_{\mathrm{lat}}^{(\nu_0)}\). 
    (c) The IQP model is then trained on high-resolution data in two stages: first, the transferred reference latent is used to learn the shared circuit core \(\theta_{\mathrm{core}}\); second, with \(\theta_{\mathrm{core}}\) fixed, the instance-dependent IQP latent variables \(\theta_{\mathrm{lat}}^{(\nu_i)}\) are sequentially adapted for new viscosities using the MMD objective. The trained IQP circuit is finally sampled and decoded through the inverse bit-to-float mapping to reconstruct high-resolution (\(64\times 64\)) solutions (Sec.~\ref{sec:physics_informed_init}).
    }
    \label{fig:scheme}
\end{figure*}

Our conceptual motivation is inspired by the Platonic Representation Hypothesis (PRH) \cite{huh2024platonic}, which posits that learning systems trained with different architectures, objectives, and even modalities may nevertheless converge toward representations of a common underlying reality. In that language, the classical PINN surrogate and the quantum IQP generative model are not viewed as solving unrelated problems, but as different parameterizations of the same underlying PDE solution manifold. This does not imply that their latent spaces are identical. Rather, it suggests that a latent representation learned by the classical surrogate may already encode useful structure for the corresponding quantum latent variable, so that the required classical-to-quantum transformation is simpler than one might expect from a fully uninformed initialization. 

In this work, we introduce a physics-informed latent initialization scheme for IQP generative learning and study its empirical consequences on parameterized Burgers' equation solution families. A latent-conditioned PINN surrogate is first trained on lower-resolution solution data, and the learned latent for a reference viscosity is transferred to initialize the corresponding IQP latent variable. The shared IQP core is then learned around this structured initialization, after which the core is fixed and the latent variables are adapted sequentially across additional viscosity values. Compared with the baseline latent-adaptation framework with random latent initialization, the proposed procedure yields more accurate reconstructions at unseen viscosities and lower MSE across all three initial-condition families considered. The improvement persists even when the PINN surrogate is trained on substantially coarser grids, suggesting that the transferred latent provides a robust physics-informed prior. Pairwise similarity analysis further shows that the classical and quantum latent spaces preserve a compatible inter-instance organization, supporting the interpretation of the transferred latent as a structured warm start for IQP adaptation.

The remainder of this paper is organized as follows. In Sec.~\ref{sec:generative_iqp}, we review the generative IQP circuit-learning framework, including the float-to-bit mapping, the MMD-based training objective, and the latent-adaptation scheme. In Sec.~\ref{sec:pinn}, we introduce the latent-conditioned PINN surrogate and its two-stage adaptation procedure for parameterized PDEs. In Sec.~\ref{sec:prh}, we discuss the PRH as the conceptual motivation for transferring latent representations across classical and quantum models. In Sec.~\ref{sec:physics_informed_init}, we formalize latent adaptation with physics-informed initialization as the main method of this work. In Sec.~\ref{sec:numerics}, we present numerical experiments on Burgers' equation, comparing physics-informed and random IQP latent initialization across multiple initial conditions, unseen viscosity values, and PINN surrogate resolutions. We also analyze the relation between the classical and quantum latent spaces through pairwise similarity preservation. In Sec.~\ref{sec:discussion}, we interpret the empirical improvement through a warm-start or effective pre-optimization perspective and discuss the broader implications and limitations of using classical physics-informed surrogate representations to initialize quantum generative models. Finally, we conclude the paper in Sec.~\ref{sec:conclusion}.

\section{Generative IQP Circuit Learning}
\label{sec:generative_iqp}

In this section, we review the generative IQP circuit learning framework that serves as the quantum backbone of our approach. The method combines three ingredients: an invertible float-to-bit representation that converts continuous data into binary strings, a train-on-classical objective based on maximum mean discrepancy (MMD), and a latent adaptation mechanism that reuses a shared IQP circuit core while adapting a low-dimensional parameter block across instances.

\subsection{Float-to-bit mapping}
\label{subsec:float_to_bit_mapping}

The output of the PDE solver is a continuous-valued field, whereas the IQP generative model produces bitstrings through measurements in the computational basis. To interface these two representations, we employ a deterministic float-to-bit mapping that converts each continuous quantity into a fixed-length binary code.

Let a scalar variable $v \in [a,b]$ lie in a bounded interval determined from the dataset. Using $N$ bits for this variable, we discretize the interval uniformly with resolution
\begin{equation}
\Delta = \frac{b-a}{2^N}.
\end{equation}
We then map $v$ to a discrete bin index
\begin{equation}
k(v) := \left\lfloor \frac{2^N(v-a)}{b-a} \right\rfloor + 1,
\qquad
k(v) \in \{1,2,\dots,2^N\}.
\end{equation}
The integer index is subsequently converted into its standard $N$-bit binary representation,
\begin{equation}
b(v) := \mathrm{BinaryEncode}_N(k(v)) \in \{0,1\}^N.
\end{equation}

For a multivariate sample, the full binary representation is obtained by concatenating the binary code of each component. In the turbulence setting of Ref.~\cite{liu2026toward}, a sample is represented by three continuous coordinates $(x,y,z)\in\mathbb{R}^3$, and the mapping is
\begin{equation}
(x,y,z)
\ \mapsto\ 
\big[\, b(x)\,\|\, b(y)\,\|\, b(z)\,\big] \in \{0,1\}^{3N},\label{eq:mapping}
\end{equation}
where $\|$ denotes concatenation. In that work, $N=6$ bits are used per coordinate, yielding an 18-bit representation and therefore an 18-qubit IQP model. In our setting, the same idea is used to map continuous PDE data into a binary representation that can be modeled by an IQP circuit.

The mapping is efficiently invertible. Hence, once bitstrings are sampled from the trained quantum model, they can be deterministically decoded back into quantized continuous values in the original physical domain. This allows the quantum generator to be trained in the bitstring domain while remaining directly connected to the target PDE solution space.

\subsection{Training Loss}
\label{subsec:training_loss}

After binarization, the target dataset is viewed as an empirical distribution over bitstrings. Let $\hat{p}_{\mathrm{data}}$ denote this empirical distribution, and let $p_{\theta}$ denote the bitstring distribution induced by an IQP circuit with parameters $\theta$.
The IQP circuit is written as
\begin{equation}
U(\theta)
=
H^{\otimes n}
\exp\!\left(
i\sum_{s\subseteq[n]} \theta_s Z_s
\right)
H^{\otimes n},
\label{eq:iqp_circuit}
\end{equation}
where $n$ is the number of qubits, $[n]=\{1,\dots,n\}$, and
\begin{equation}
Z_s := \bigotimes_{i\in s} Z_i
\end{equation}
denotes the multi-qubit Pauli-$Z$ string supported on subset $s$. Sampling in the computational basis defines the model distribution
\begin{equation}
x \sim q_{\theta}(x) := \left|\langle x|U(\theta)|0^n\rangle\right|^2.
\end{equation}

To train the model, we minimize the maximum mean discrepancy (MMD) between the model distribution and the empirical data distribution. The loss is
\begin{eqnarray}
&&\mathcal{L}_{\mathrm{MMD}}(\theta)
=
\mathbb{E}_{b,b'\sim p_{\theta}}\!\left[k(b,b')\right] \nonumber\\
&&-2\mathbb{E}_{b\sim p_{\theta},\,b'\sim \hat{p}_{\mathrm{data}}}\!\left[k(b,b')\right]
+\mathbb{E}_{b,b'\sim \hat{p}_{\mathrm{data}}}\!\left[k(b,b')\right],
\label{eq:mmd_loss}
\end{eqnarray}
where $k(b,b')$ is a positive kernel on bitstrings, taken in Ref.~\cite{liu2026toward, recio2025train} to be a Gaussian kernel.

A key advantage of the train-on-classical, deploy-on-quantum paradigm is that for IQP circuits this loss can be rewritten in terms of expectation values of commuting Pauli observables:
\begin{equation}
\mathcal{L}_{\mathrm{MMD}}(\theta)
=
\sum_{s\subseteq[n]} c_s
\left(
\langle Z_s\rangle_{\theta}
-
\langle Z_s\rangle_{\mathrm{data}}
\right)^2,
\label{eq:mmd_pauli}
\end{equation}
where the coefficients $c_s$ depend only on the chosen kernel, $\langle Z_s\rangle_{\theta}$ is the expectation value under the IQP model, and $\langle Z_s\rangle_{\mathrm{data}}$ is the corresponding empirical moment of the dataset.

Because these quantities are classically tractable for IQP circuits, the full optimization loop can be carried out on classical hardware:
\begin{equation}
\theta^\ast
=
\arg\min_{\theta}\mathcal{L}_{\mathrm{MMD}}(\theta).
\end{equation}
After optimization, the learned parameters $\theta^\ast$ can be loaded onto a quantum device for sampling. The generated bitstrings are then decoded through the inverse float-to-bit mapping to recover samples in the original continuous domain.

\subsection{Latent Adaptation}
\label{subsec:latent_adaptation}

A central idea in the recent proposed generative IQP framework \cite{liu2026toward} is that one does not retrain an entirely new circuit for every related instance. Instead, one reuses a shared circuit core and adapts only a low-dimensional latent block. This is particularly suitable when the target data correspond to the same physical system under different parameter values.

Let $\mathcal{D}_t=\{x_i^{(t)}\}_{i=1}^{N_t}$ denote the binarized dataset associated with instance $t$, with empirical distribution $\hat{p}_t$. The IQP generator is still given by Eq.~\eqref{eq:iqp_circuit}, but the parameter vector is partitioned as
\begin{equation}
\theta = (\theta_{\mathrm{core}}, \theta_{\mathrm{lat}}),
\qquad
\dim(\theta_{\mathrm{lat}})=d_{\mathrm{lat}}\ll \dim(\theta_{\mathrm{core}}).
\label{eq:param_partition}
\end{equation}
Here, $\theta_{\mathrm{core}}$ contains the shared parameters of the circuit, while $\theta_{\mathrm{lat}}$ is a low-dimensional instance-dependent latent block.

For the first training instance, one initializes the latent block \textit{randomly} and keeps it fixed, while optimizing only the shared core:
\begin{equation}
\theta_{\mathrm{core}}^{(1)}
\in
\arg\min_{\theta_{\mathrm{core}}}
\mathcal{L}_{\mathrm{MMD}}
\bigl(
q_{(\theta_{\mathrm{core}},\,\theta_{\mathrm{lat}}^{(1)})},
\hat{p}_1
\bigr).
\label{eq:first_stage_core}
\end{equation}
After this initial fit, the learned core parameters are frozen and reused for all subsequent instances,
\begin{equation}
\theta_{\mathrm{core}} \leftarrow \theta_{\mathrm{core}}^{(1)}.
\end{equation}
Then, for each new instance $t\ge 2$, only the latent block is adapted:
\begin{equation}
\theta_{\mathrm{lat}}^{(t)}
\in
\arg\min_{\theta_{\mathrm{lat}}}
\mathcal{L}_{\mathrm{MMD}}
\bigl(
q_{(\theta_{\mathrm{core}}^{(1)},\,\theta_{\mathrm{lat}})},
\hat{p}_t
\bigr),
\label{eq:latent_adapt}
\end{equation}
with the optimization initialized from the previous solution $\theta_{\mathrm{lat}}^{(t-1)}$.

This warm-start strategy assumes continuity across neighboring instances: if adjacent parameter values correspond to similar target distributions, then only a small update in the latent block is needed to track the change, while the shared core continues to encode the dominant common structure. A geometric motivation for this latent-adaptation assumption is provided in Appendix~\ref{app:latent_adaptation_motivation}.

The adapted latent parameters
\begin{equation}
\left\{\theta_{\mathrm{lat}}^{(t)}\right\}_{t=1}^{T}
\end{equation}
therefore define a low-dimensional trajectory in parameter space,
\begin{equation}
t \mapsto \theta_{\mathrm{lat}}^{(t)} \in \mathbb{R}^{d_{\mathrm{lat}}}.
\end{equation}
To generate an unseen instance at an intermediate value $\tau$, one interpolates between nearby latent anchors. For piecewise-linear interpolation, if $k=\lfloor \tau \rfloor$ and $\alpha=\tau-k\in[0,1]$, then
\begin{equation}
\tilde{\theta}_{\mathrm{lat}}(\tau)
=
(1-\alpha)\theta_{\mathrm{lat}}^{(k)}
+
\alpha \theta_{\mathrm{lat}}^{(k+1)}.
\label{eq:latent_interp}
\end{equation}
More generally, one may write
\begin{equation}
\tilde{\theta}_{\mathrm{lat}}(\tau)
=
\sum_{t=1}^{T} w_t(\tau)\theta_{\mathrm{lat}}^{(t)},
\qquad
\sum_{t=1}^{T} w_t(\tau)=1,
\end{equation}
with suitable interpolation, or even extrapolation, weights.

The resulting generator for an unseen instance is thus
\begin{equation}
x \sim q_{(\theta_{\mathrm{core}}^{(1)},\,\tilde{\theta}_{\mathrm{lat}}(\tau))}(x),
\end{equation}
followed by inverse decoding back to the continuous domain. In this way, a single compact IQP circuit, together with a learned latent trajectory, can represent an entire family of related solution distributions, under the assumption that the underlying distributions evolve smoothly across adjacent parameter values.

In the baseline latent-adaptation framework, the latent variable for the first instance is initialized randomly. Our contribution in later sections is to replace this random initialization with a physics-informed initialization obtained from a classical surrogate model, thereby injecting task-structured prior information into the latent adaptation process.

\section{Physics-Informed Neural Network Methods}
\label{sec:pinn}

To construct a physics-informed classical surrogate for parameterized PDE solution families, we employ a PINN. The role of this surrogate is twofold. First, it provides a continuous parametric model of the solution field that is constrained both by observed data and by the governing differential equation. Second, by endowing each problem instance with its own low-dimensional latent variable, it yields a latent representation that can be transferred to initialize the quantum latent adaptation procedure introduced in Sec.~\ref{sec:generative_iqp} as will be explored in the following sections. In this work, we instantiate this framework using Burgers' equation as a representative and elementary PDE benchmark.

\subsection{General PINN formulation}
\label{subsec:general_pinn}

Consider a partial differential equation of the form
\begin{equation}
\mathcal{N}[u](x,t;\nu)=0,
\label{eq:general_pde}
\end{equation}
where \(u(x,t;\nu)\) denotes the unknown solution field and \(\nu\) is the physical parameter of interest, and \(\mathcal{N}\) is a differential operator encoding the governing PDE. Using Burgers' equation as an example, the governing PDE is
\begin{equation}
\frac{\partial u}{\partial t}
+
u\,\frac{\partial u}{\partial x}
-
\nu\,\frac{\partial^2 u}{\partial x^2}
= 0,
\label{eq:burgers_eq}
\end{equation}
with $\nu$ the viscosity of the system.
In a standard PINN, one introduces a neural-network ansatz
\begin{equation}
u_{\Theta}(x,t) \approx u(x,t),
\end{equation}
where \(\Theta\) denotes the trainable parameters of the classical neural network. The network is trained not only to fit observed solution values, but also to satisfy the governing PDE through automatic differentiation. The training objective is therefore composed of a data-fitting term and a physics term,
\begin{equation}
\mathcal{L}_{\mathrm{PINN}}(\Theta)
=
\lambda_{\mathrm{data}}\,\mathcal{L}_{\mathrm{data}}(\Theta)
+
\lambda_{\mathrm{phys}}\,\mathcal{L}_{\mathrm{phys}}(\Theta),
\label{eq:pinn_total_loss}
\end{equation}
where \(\lambda_{\mathrm{data}},\lambda_{\mathrm{phys}}>0\) balance the two contributions.

Given supervised samples
\begin{equation}
\mathcal{D}^{(\nu)}=\{(x_i,t_i,u_i)\}_{i=1}^{N_\nu},
\end{equation}
the data loss is taken as the mean-squared error
\begin{equation}
\mathcal{L}_{\mathrm{data}}(\Theta)
=
\frac{1}{N_\nu}\sum_{i=1}^{N_\nu}
\left|
u_{\Theta}(x_i,t_i)-u_i
\right|^2.
\label{eq:data_loss_standard}
\end{equation}

The physics loss is defined by evaluating the PDE residual on collocation points. For Burgers' equation, the residual associated with \(u_{\Theta}\) is
\begin{eqnarray}
&&r_{\Theta}(x,t;\nu)
=
\partial_t u_{\Theta}(x,t)
+
u_{\Theta}(x,t)\,\partial_x u_{\Theta}(x,t) \nonumber \\
&&-
\nu\,\partial_{xx}u_{\Theta}(x,t),
\label{eq:burgers_residual_standard}
\end{eqnarray}
and the corresponding physics loss is
\begin{equation}
\mathcal{L}_{\mathrm{phys}}(\Theta)
=
\frac{1}{N_r}\sum_{j=1}^{N_r}
\left|
r_{\Theta}(\tilde{x}_j,\tilde{t}_j;\nu)
\right|^2.
\label{eq:physics_loss_standard}
\end{equation}
In practice, the derivatives in Eq.~\eqref{eq:burgers_residual_standard} are computed by automatic differentiation in \texttt{Pytorch}.

\subsection{Latent-conditioned PINN surrogate}
\label{subsec:latent_pinn}

To parallel the quantum latent-adaptation picture, we do not use a separate fully independent PINN for each viscosity. Instead, we introduce a \emph{shared} neural generator together with an \emph{instance-dependent latent vector}. Concretely, the network takes spatial-temporal coordinates \((x,t)\) together with a latent code \(z\in\mathbb{R}^{d_z}\) and outputs the corresponding solution value,
\begin{equation}
u_\Theta(x,t;\,z) = G_\Theta(x,t,z),
\label{eq:latent_generator}
\end{equation}
where \(G_\Theta\) is implemented as a multilayer perceptron (MLP). The input to the MLP is the concatenated vector
\begin{equation}
\bigl[x,\;t,\;z\bigr]\in\mathbb{R}^{2+d_z},
\end{equation}
and the network outputs a scalar prediction \(u\). Thus, for each viscosity instance \(\nu\), the latent vector \(z^{(\nu)}\) serves as a compact representation of that specific solution manifold, while the network parameters \(\Theta\) are shared across all instances.

Accordingly, for each viscosity \(\nu\), the model prediction is written as
\begin{equation}
u^{(\nu)}_\Theta(x,t)
=
G_\Theta\!\left(x,t,z^{(\nu)}\right).
\label{eq:instance_prediction}
\end{equation}
This is the classical analogue of the shared-core plus latent-block parameterization used in the IQP model. 

\subsection{Data and physics losses with latent variables}
\label{subsec:pinn_losses_latent}

Given a collection of viscosities \(\{\nu_m\}_{m=1}^{M}\), we assign one latent vector \(z^{(\nu_m)}\) to each instance and jointly optimize the shared network parameters \(\Theta\) together with all latent vectors. For a single viscosity \(\nu\), the data loss is
\begin{equation}
\mathcal{L}_{\mathrm{data}}^{(\nu)}(\Theta,z^{(\nu)})
=
\frac{1}{N_\nu}\sum_{i=1}^{N_\nu}
\left|
G_\Theta(x_i,t_i,z^{(\nu)}) - u_i
\right|^2.
\label{eq:data_loss_latent}
\end{equation}

The Burgers residual associated with the latent-conditioned model is
\begin{eqnarray}
&&r_{\Theta}^{(\nu)}(x,t;z^{(\nu)})
=
\partial_t G_\Theta(x,t,z^{(\nu)}) -
\nu\,\partial_{xx}G_\Theta(x,t,z^{(\nu)}) \nonumber \\
&&+
G_\Theta(x,t,z^{(\nu)})\,\partial_x G_\Theta(x,t,z^{(\nu)})
,
\label{eq:latent_burgers_residual}
\end{eqnarray}
and the corresponding physics loss is
\begin{equation}
\mathcal{L}_{\mathrm{phys}}^{(\nu)}(\Theta,z^{(\nu)})
=
\frac{1}{N_r^{(\nu)}}\sum_{j=1}^{N_r^{(\nu)}}
\left|
r_{\Theta}^{(\nu)}(\tilde{x}_j,\tilde{t}_j;z^{(\nu)})
\right|^2.
\label{eq:physics_loss_latent}
\end{equation}
The overall objective is then
\begin{eqnarray}
&&\mathcal{L}_{\mathrm{PINN}}(\Theta,\{z^{(\nu)}\}) \nonumber \\
&&=
\sum_{\nu\in\mathcal{V}}
\left[
\lambda_{\mathrm{data}}\,
\mathcal{L}_{\mathrm{data}}^{(\nu)}(\Theta,z^{(\nu)})
+
\lambda_{\mathrm{phys}}\,
\mathcal{L}_{\mathrm{phys}}^{(\nu)}(\Theta,z^{(\nu)})
\right]. \nonumber \\
\label{eq:joint_pinn_loss}
\end{eqnarray}

In the implementation, the data term is taken as the mean-squared error between predicted and target solution values, while the physics term is the mean-squared PDE residual computed through automatic differentiation. The resulting optimization is then organized into two stages, in direct analogy with the IQP latent-adaptation scheme. First, the shared surrogate parameters and the latent variable associated with a reference viscosity are learned jointly. Subsequently, the shared surrogate is held fixed, and only the latent variable is adapted for new viscosity instances.

\subsection{Stage-I training: learning a shared surrogate and latent codes}
\label{subsec:stageA}

The first training stage learns the shared classical surrogate \(G_{\Theta}\) using a single reference viscosity \(\nu_0\), in direct analogy with the learning of \(\theta_{\mathrm{core}}\) in the IQP latent-adaptation scheme. At this stage, we introduce one trainable latent vector \(z^{(\nu_0)}\) for the reference instance and jointly optimize it together with the shared network parameters. Specifically, Stage-I solves
\begin{eqnarray}
&&(\Theta^\ast, z^{(\nu_0)\ast}) 
\in
\arg\min_{\Theta,\, z^{(\nu_0)}}
[
\lambda_{\mathrm{data}}\,
\mathcal{L}_{\mathrm{data}}^{(\nu_0)}(\Theta,z^{(\nu_0)}) \nonumber \\
&&+
\lambda_{\mathrm{phys}}\,
\mathcal{L}_{\mathrm{phys}}^{(\nu_0)}(\Theta,z^{(\nu_0)})
].
\label{eq:stageA_single}
\end{eqnarray}
After this initial fit, the shared network parameters \(\Theta^\ast\) are frozen and reused for all subsequent viscosity instances, while only the latent variable is adapted. In this way, \(\Theta^\ast\) plays the role of a shared classical surrogate core, and \(z^{(\nu)}\) captures the instance-specific variation across viscosities.

\begin{figure*}[t]
    \includegraphics[width=0.8\textwidth]{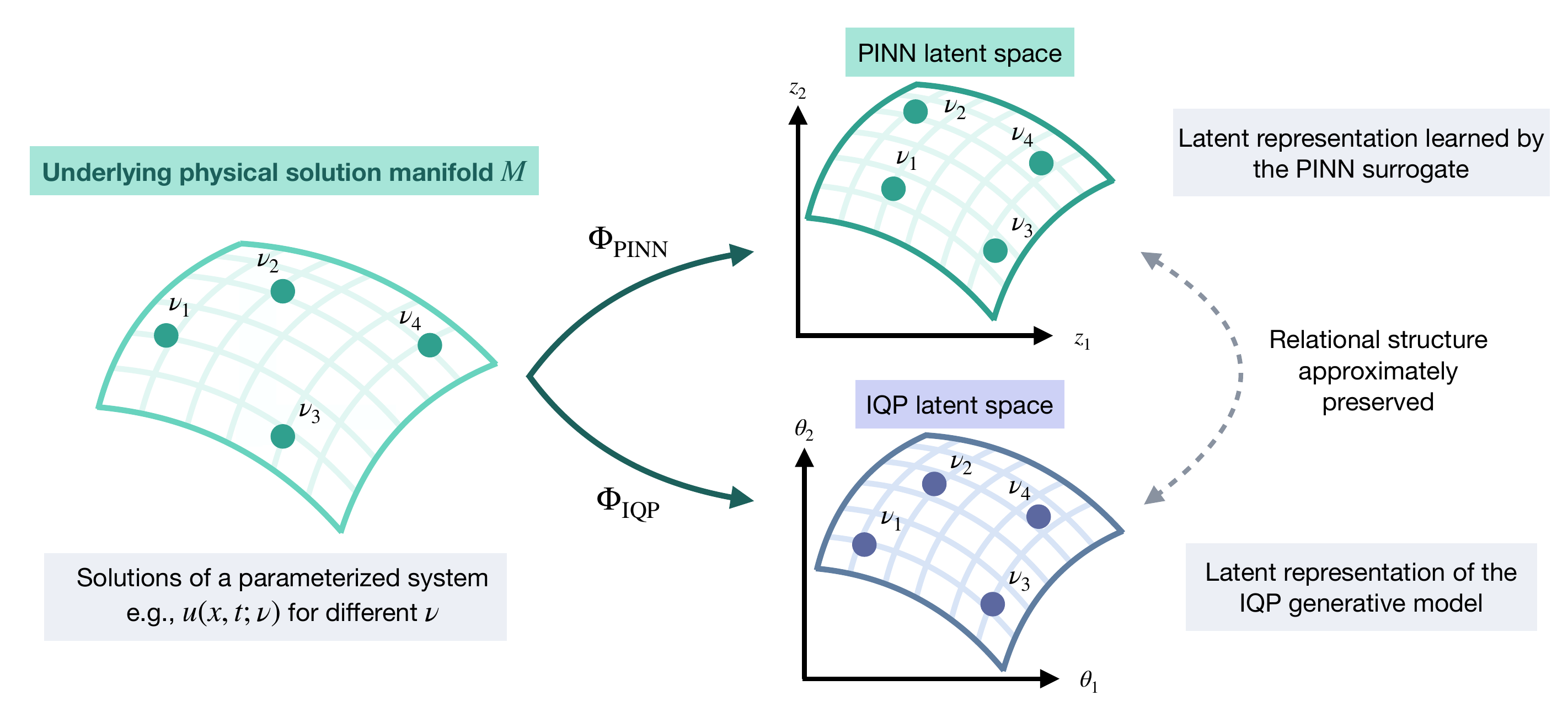}
    \caption{
    PRH-motivated view of latent alignment between the PINN surrogate and the IQP generative model. 
    A parameterized family of physical solutions forms an underlying solution manifold \(\mathcal{M}\). 
    The PINN and IQP models induce different latent representations of this same family through \(\Phi_{\rm PINN}\) and \(\Phi_{\rm IQP}\). 
    While the two latent coordinate systems need not match directly, their inter-instance relational structure may be approximately preserved. 
    This motivates transferring the PINN latent representation to initialize the IQP latent variable.
    }
    \label{fig:prh_latent_alignment}
\end{figure*}

\subsection{Stage-II latent adaptation for new viscosities}
\label{subsec:stageB}

After the shared surrogate has been trained, we freeze its parameters and adapt only the latent variable for a new viscosity value. Given a new instance \(\nu_{\mathrm{new}}\), with corresponding supervised samples \(\mathcal{D}^{(\nu_{\mathrm{new}})}\), Stage-II solves
\begin{eqnarray}
&&z^{(\nu_{\mathrm{new}})\ast}
\in
\arg\min_{z}
[
\lambda_{\mathrm{data}}\,
\mathcal{L}_{\mathrm{data}}^{(\nu_{\mathrm{new}})}(\Theta^\ast,z) \nonumber \\
&& +
\lambda_{\mathrm{phys}}\,
\mathcal{L}_{\mathrm{phys}}^{(\nu_{\mathrm{new}})}(\Theta^\ast,z)
].
\label{eq:stageB}
\end{eqnarray}
That is, the surrogate network remains fixed, and only the low-dimensional latent representation is adapted for the new problem instance.

This mirrors the IQP latent-adaptation protocol, in which the shared circuit core is fixed and only the latent parameter block is optimized for a new instance. In the PINN setting, this means that the shared neural operator-like backbone is reused across viscosities, while the instance-dependent latent vector absorbs the change in the physical parameter.

\subsection{Sequential latent adaptation across viscosity}
\label{subsec:sequential_latent}

In addition to adapting a single new instance, it is also possible to perform sequential latent adaptation across a viscosity sweep. Starting from an initial latent vector \(z^{(\nu_0)}\), one optimizes the latent code for a sequence of viscosities
$\nu_0,\nu_1,\dots,\nu_K$,
where each new optimization is initialized from the previously learned latent:
\begin{equation}
z^{(\nu_{k})}_{\mathrm{init}} = z^{(\nu_{k-1})\ast}.
\label{eq:warm_start_latent}
\end{equation}
For each \(k\ge 1\), one then solves
\begin{equation}
z^{(\nu_k)\ast}
\in
\arg\min_{z}
\left[
\lambda_{\mathrm{data}}\,
\mathcal{L}_{\mathrm{data}}^{(\nu_k)}(\Theta^\ast,z)
+
\lambda_{\mathrm{phys}}\,
\mathcal{L}_{\mathrm{phys}}^{(\nu_k)}(\Theta^\ast,z)
\right].
\label{eq:sequential_adapt}
\end{equation}

This warm-start procedure is based on the assumption that nearby viscosity values induce nearby solution distributions, so that their latent representations should also vary smoothly. The resulting sequence
\begin{equation}
\left\{z^{(\nu_k)\ast}\right\}_{k=0}^{K}
\end{equation}
defines a classical latent trajectory across parameter space.

\section{The Platonic Representation Hypothesis}
\label{sec:prh}

Our motivation for transferring latent variables from the classical PINN surrogate to the quantum IQP model is inspired by the Platonic Representation Hypothesis (PRH) \cite{huh2024platonic}. At a conceptual level, as illustrated in Fig.~\ref{fig:prh_latent_alignment}, PRH posits that distinct learning systems, even when trained with different architectures, objectives, or data modalities, may converge toward representations of a common underlying reality. In that view, observed data are different projections of an underlying latent structure, and increasingly capable models tend to organize their internal representations according to that shared structure.

Adapting this viewpoint to the present setting, we regard the solution family of a parameterized PDE as the underlying physical object of interest. The classical PINN surrogate and the quantum IQP generative model are then two different parameterizations intended to describe the same solution manifold. Although their latent coordinates are not expected to coincide exactly, we suppose that both are shaped by the same underlying physical structure.
More formally, let
\begin{equation}
\mathcal{M}
=
\left\{
u(\cdot,\cdot;\nu)\,:\,\nu\in\mathcal{V}
\right\}
\end{equation}
denote the family of PDE solutions over the parameter domain \(\mathcal{V}\). We then view the classical and quantum models as inducing two representation maps of the same solution family,
\begin{equation}
\Phi_{\mathrm{PINN}}:\mathcal{M}\to\mathbb{R}^{d_c},
\qquad
\Phi_{\mathrm{IQP}}:\mathcal{M}\to\mathbb{R}^{d_q},
\end{equation}
where \(\Phi_{\mathrm{PINN}}(u(\cdot,\cdot;\nu)) = z_{\mathrm{PINN}}^{(\nu)}\) denotes the latent representation learned by the classical surrogate, and \(\Phi_{\mathrm{IQP}}(u(\cdot,\cdot;\nu))\) corresponds to the latent variable required for IQP adaptation.

The PRH-motivated assumption is not that these two representations are identical, nor that there exists a known explicit map from one latent space to the other. Rather, it is that they may preserve a related instance-wise organization of the same underlying solution manifold. Concretely, for PDE instances $u_i,u_j \in \mathcal{M}$, we assume that the relational structure induced by the quantum and classical representations is approximately aligned in the sense that
\begin{equation}
\operatorname{Rel}\!\left(\Phi_{\mathrm{IQP}}(u_i),\Phi_{\mathrm{IQP}}(u_j)\right)
\approx
\operatorname{Rel}\!\left(\Phi_{\mathrm{PINN}}(u_i),\Phi_{\mathrm{PINN}}(u_j)\right),
\label{eq:prh_rel}
\end{equation}
where $\operatorname{Rel}(\cdot,\cdot)$ denotes a generic relation measure, such as neighborhood, similarity, or distance in latent space. In the most optimistic case, this means that nearby PDE instances in the PINN latent space remain nearby in the corresponding IQP latent space, even if the coordinates themselves are not directly matched. Under this interpretation, the latent representation learned by the PINN may still provide a meaningful structured initialization for the corresponding IQP latent variable, which we formalize in the next section.

\section{Latent Adaptation with Physics-Informed Initialization}
\label{sec:physics_informed_init}

We now formalize the main method of this work. The central idea is to replace the random initialization used in standard IQP latent adaptation with a physics-informed initialization obtained from the classical PINN surrogate. More specifically, the PINN latent learned from the lower-resolution solution space is transferred as the initialization of the IQP latent variable for the corresponding higher-resolution instance. For each parameter value \(\nu\), let
\begin{equation}
\mathcal{D}_{\nu}=\{x_i^{(\nu)}\}_{i=1}^{N_\nu}
\end{equation}
denote the corresponding binarized dataset, with empirical distribution \(\hat{p}_{\nu}\).
For the reference viscosity \(\nu_0\), we initialize
\begin{equation}
z_{\mathrm{PINN}}^{(\nu_0)}
\;\xrightarrow{\mathrm{init}}\;
\theta_{\mathrm{lat}}^{(\nu_0)}.
\label{eq:ref_init}
\end{equation}
This initialized latent is then held fixed while the shared IQP core is learned by solving
\begin{equation}
\theta_{\mathrm{core}}
\in
\arg\min_{\theta_{\mathrm{core}}}
\mathcal{L}_{\mathrm{MMD}}
\!\left(
q_{(\theta_{\mathrm{core}},\,\theta_{\mathrm{lat}}^{(\nu_0)})},
\hat{p}_{\nu_0}
\right),
\label{eq:core_train_fixed_latent}
\end{equation}
where \(q_{(\theta_{\mathrm{core}},\,\theta_{\mathrm{lat}})}\) denotes the IQP-generated bitstring distribution induced by the circuit parameters \((\theta_{\mathrm{core}},\theta_{\mathrm{lat}})\), as introduced in Sec.~\ref{subsec:training_loss}.
After the shared core \(\theta_{\mathrm{core}}\) has been learned from the reference instance, latent adaptation for a new parameter value \(\nu\) is performed with \(\theta_{\mathrm{core}}\) fixed, using the previously adapted latent as the warm start.

For the first adaptation step, we take \(\nu_{\mathrm{prev}}=\nu_0\). For a new viscosity \(\nu\), let \(\theta_{\mathrm{lat},0}^{(\nu)}\) denote the initial latent value used for adaptation, which is set equal to the previously adapted latent,
\[
\theta_{\mathrm{lat},0}^{(\nu)}
=
\theta_{\mathrm{lat}}^{(\nu_{\mathrm{prev}})}.
\]
The sequential adaptation procedure is therefore summarized as
\begin{equation}
\theta_{\mathrm{lat}}^{(\nu_{\mathrm{prev}})}
\;\xrightarrow{\mathrm{warm\ start}}\;
\theta_{\mathrm{lat},0}^{(\nu)}
\;\xrightarrow[\theta_{\mathrm{core}}\ \mathrm{fixed}]{\mathrm{adapt}}
\theta_{\mathrm{lat}}^{(\nu)}.
\label{eq:sequential_adapt}
\end{equation}

More explicitly, \(\theta_{\mathrm{lat}}^{(\nu)}\) is obtained by solving
\begin{equation}
\theta_{\mathrm{lat}}^{(\nu)}
\in
\arg\min_{\theta_{\mathrm{lat}}}
\mathcal{L}_{\mathrm{MMD}}
\!\left(
q_{(\theta_{\mathrm{core}},\theta_{\mathrm{lat}})},
\hat{p}_{\nu}
\right).
\label{eq:latent_adapt_physics_init}
\end{equation}
Proceeding in the same manner as in Sec.~\ref{subsec:latent_adaptation}, the adapted latent parameters
\begin{equation}
\left\{\theta_{\mathrm{lat}}^{(\nu_k)}\right\}_{k=0}^{K}
\end{equation}
define a low-dimensional trajectory in parameter space,
\begin{equation}
\nu \mapsto \theta_{\mathrm{lat}}^{(\nu)}
\in
\mathbb{R}^{d_{\mathrm{lat}}}.
\end{equation}
Interpolation between nearby latent anchors may then be carried out exactly as in Eq.~\eqref{eq:latent_interp}, yielding an interpolated latent \(\tilde{\theta}_{\mathrm{lat}}(\tilde{\nu})\) for an unseen intermediate parameter value \(\tilde{\nu}\). The corresponding generator is
\begin{equation}
x \sim q_{(\theta_{\mathrm{core}},\,\tilde{\theta}_{\mathrm{lat}}(\tilde{\nu}))}(x),
\end{equation}
followed by inverse decoding back to the continuous domain. In this way, a single compact IQP circuit, together with a learned latent trajectory, can represent an entire family of related solution distributions, provided that the underlying distributions vary smoothly across adjacent parameter values. The overall proposed scheme is illustrated in Fig.~\ref{fig:scheme}.

\begin{figure}[t]
    \includegraphics[width=0.45\textwidth]{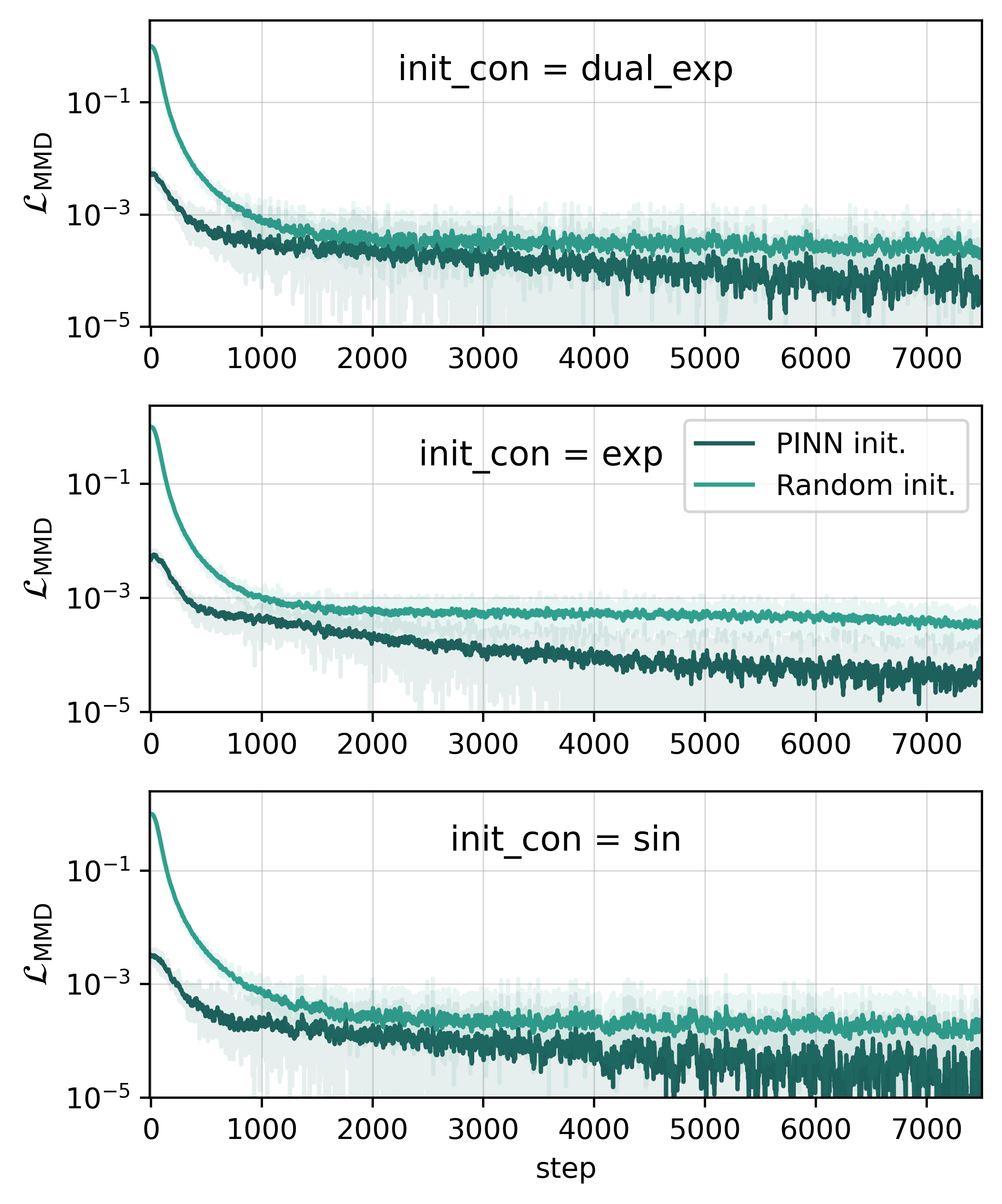}
    \caption{
    Training loss during the reference-core learning step at \(\nu_0=0.06\), comparing physics-informed initialization and random initialization of the IQP latent variable. The three panels correspond to the Burgers' equation initial conditions \texttt{dual\_exp}, \texttt{exp}, and \texttt{sin}. In all cases, the PINN-initialized run uses a latent representation transferred from a \(32\times32\) classical surrogate. The plotted quantity is the MMD objective used to train the shared IQP core \(\theta_{\mathrm{core}}\).
    }
    \label{fig:training_loss_core}
\end{figure}

\begin{figure*}[t]
    \includegraphics[width=0.8\textwidth]{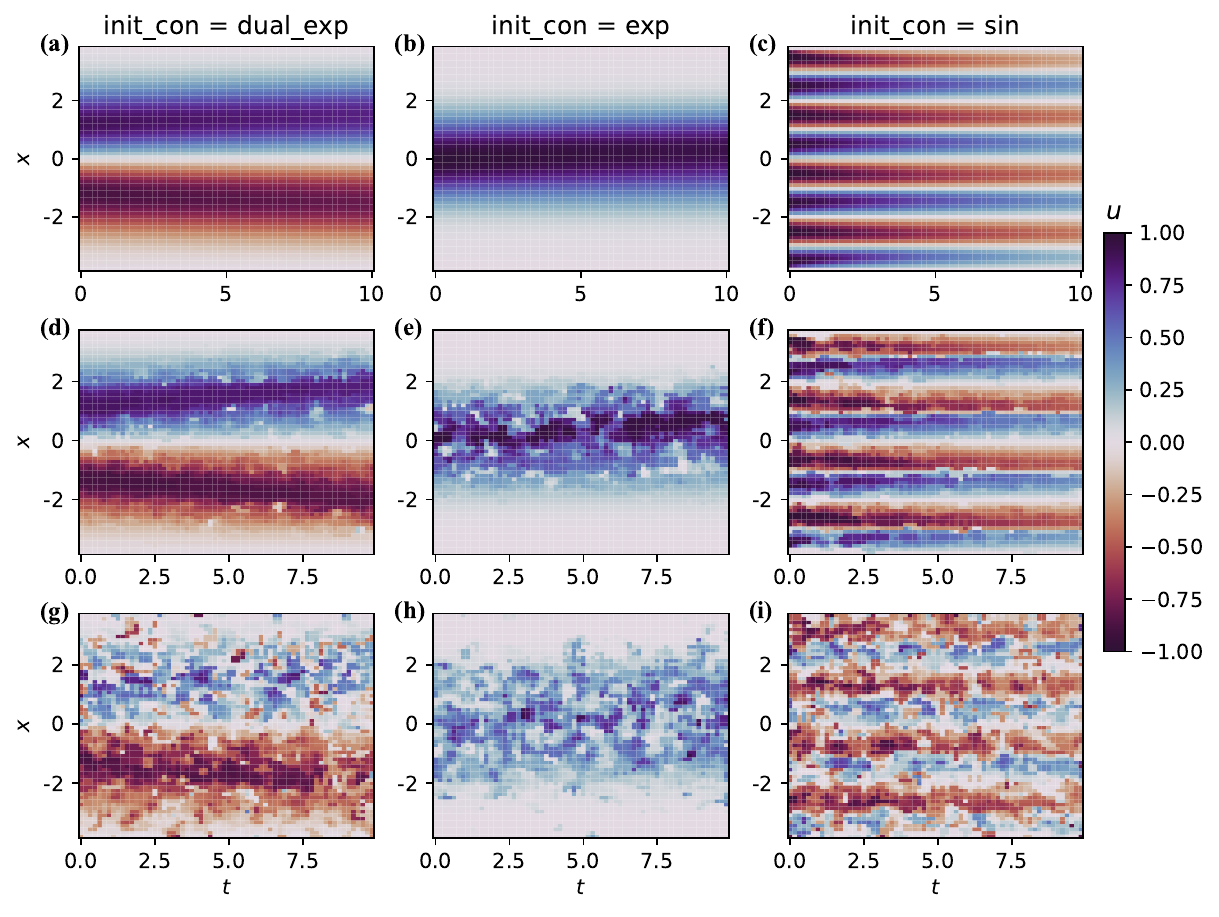}
    \caption{
    Qualitative comparison between physics-informed latent initialization and random latent initialization at the unseen viscosity value \(\nu=0.28\). The IQP prediction is obtained by interpolating between neighboring latent anchors and reconstructing the solution from \(10^6\) measurement shots. Columns correspond to the three initial conditions \texttt{dual\_exp}, \texttt{exp}, and \texttt{sin}. The top row panels (a)--(c) show the numerical ground-truth solutions of Burgers' equation, the middle row panels (d)--(f) show the predictions obtained with physics-informed initialization of the IQP latent adaptation, and the bottom row panels (g)--(i) shows the corresponding predictions from the baseline latent-adaptation scheme with random initialization. 
    }
    \label{fig:general_improvement}
\end{figure*}

\section{Numerical Experiments}
\label{sec:numerics}

\begin{figure*}[t]
    \includegraphics[width=0.95\textwidth]{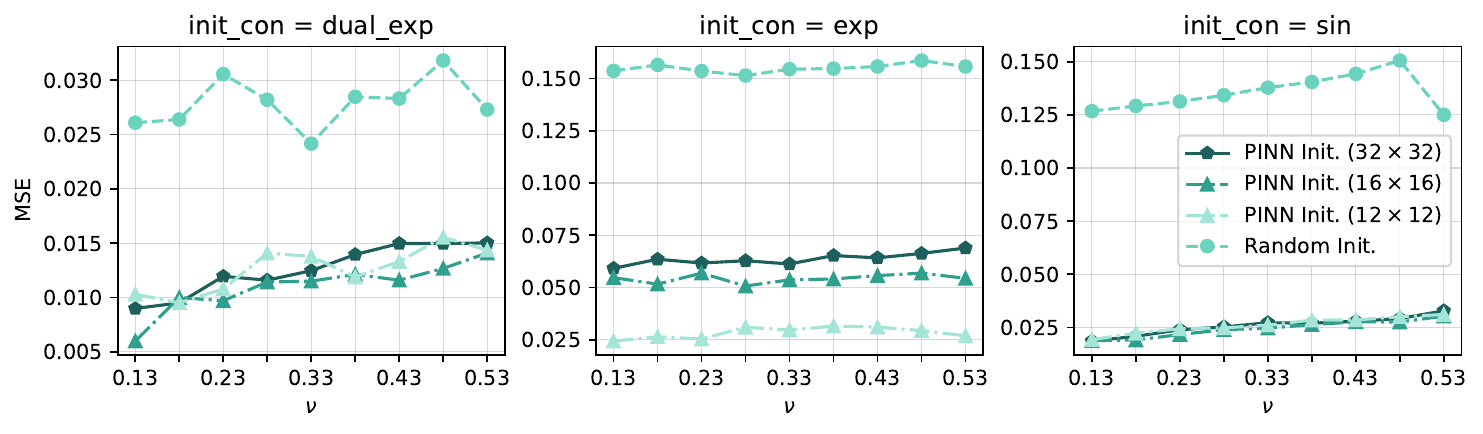}
    \caption{
    Mean-squared error (MSE) between the reconstructed IQP predictions and the numerical solutions across unseen viscosity values \(\nu \in \{0.13, 0.18, 0.23, 0.28, 0.33, 0.38, 0.43, 0.48, 0.53\}\). The three panels correspond to the initial conditions \texttt{dual\_exp}, \texttt{exp}, and \texttt{sin}. We compare the baseline latent-adaptation method with random initialization against physics-informed initialization obtained from classical PINN surrogates trained at resolutions \(32\times32\), \(16\times16\), and \(12\times12\). The PINN-initialized variants consistently outperform the random-initialization baseline, while the dependence on surrogate resolution is not entirely uniform across initial conditions; in particular, for the \texttt{exp} case, the \(12\times12\) surrogate yields the lowest MSE among the tested resolutions.
    }
    \label{fig:different_low_res}
\end{figure*}

\subsection{Experimental setup}
\label{subsec:exp_setup}

We use Burgers' equation, introduced in Eq.~\eqref{eq:burgers_eq}, as the benchmark PDE system for all numerical experiments. The solution domain is taken to be
\begin{equation}
x \in (-3.8,\,3.8),
\qquad
t \in (0,\,10),
\end{equation}
and we consider a family of solution instances parameterized by the viscosity \(\nu\).
To test the robustness of the proposed method across qualitatively different solution families, we consider the three initial conditions listed in Table~\ref{tab:init_conditions}.
\begin{table}[h]
\centering
\caption{Initial conditions used in the Burgers' equation experiments.}
\label{tab:init_conditions}
\begin{tabular}{|l|c|}
\hline
Label & Initial condition \(u(x,0)\) \\
\hline
\texttt{dual\_exp} &
\(\displaystyle
\exp\!\left(-\frac{(x-1)^2}{2}\right)
-
\exp\!\left(-\frac{(x+1)^2}{2}\right)
\) \\[1.2ex]
\texttt{exp} &
\(\displaystyle
\exp\!\left(-\frac{x^2}{2}\right)
\) \\[1.2ex]
\texttt{sin} &
\(\displaystyle
\sin(\pi x)
\) \\
\hline
\end{tabular}
\end{table}
For the baseline quantum latent-adaptation method, we generate numerical solutions of Burgers' equation on a \(64\times 64\) space-time $(x,t)$ grid for the training set of viscosity values
\begin{eqnarray}
&&\nu \in \{0.06,\,0.11,\,0.16,\,0.21,\, \nonumber \\
&&0.26,\,0.31,\,0.36,\,0.41,\,0.46,\,0.51\}.
\end{eqnarray}
Following the latent-adaptation framework of Sec.~\ref{subsec:latent_adaptation}, we train the shared IQP core using the reference viscosity \(\nu_0=0.06\), with the baseline setting associated with latent variable randomly initialized and then held fixed while learning \(\theta_{\mathrm{core}}\). Latent adaptation is subsequently performed for the remaining viscosity values in the training set.

The latent dimension is fixed to $d_{\mathrm{lat}} = 50$, following~\cite{liu2026toward},
meaning that the first 50 trainable parameters of the IQP circuit are assigned to the latent block, where $\dim(\theta_{\mathrm{core}}) \approx 63\text{k}$ thus $d_{\mathrm{lat}}\ll \dim(\theta_{\mathrm{core}})$. In the present implementation, these correspond to lower-order parameterized Pauli rotations, beginning with single-qubit terms \(\{0\}, \{1\}, \ldots\), followed by lower-order multi-qubit terms such as \(\{0,1\}, \{0,2\}, \ldots\). For the float-to-bit encoding, we use \(N=6\) bits per coordinate \((x,t,u)\). Consequently, the IQP generative model uses $3N = 18$
qubits in total. The IQP circuit includes parameterized Pauli-string gates up to locality 7, i.e., up to 7-qubit Pauli-\(Z\) strings in the diagonal unitary.

All IQP optimizations are performed with the Adam optimizer, using learning rate \(10^{-4}\) and \(7500\) iterations for each viscosity instance. As a generative modeling task, the trained IQP circuit for each viscosity is used to produce samples over the full solution domain in the encoded \((x,t,u)\) space with $10^5-10^6$ shots. At resolution \(64\times64\times64\), the first two axes correspond to the discretized spatial and temporal coordinates, while the third corresponds to the discretized solution value \(u\).

The physics-informed variant differs from the baseline only in the initialization of the reference latent variable. Instead of random initialization at \(\nu_0=0.06\), we initialize the IQP latent using the latent representation obtained from a classical PINN surrogate at the same viscosity. This surrogate is trained on a lower-resolution solution space, typically \(32\times 32\), and in additional experiments also on \(16\times16\) and \(12\times12\) grids. The latent used for \(\nu_0=0.06\) in the PINN surrogate is itself obtained through classical latent adaptation (Sec.~\ref{subsec:sequential_latent}) starting from \(\nu=0.01\).

To examine the effect of initialization on the optimization dynamics of the reference-stage training, we also record the MMD loss during the learning of the shared IQP core \(\theta_{\mathrm{core}}\) at \(\nu_0=0.06\). Fig.~\ref{fig:training_loss_core} compares the resulting loss trajectories for physics-informed initialization and random initialization under identical optimization settings, for the three initial-condition families considered in this work.

After training \(\theta_{\mathrm{core}}\) and adapting the latent variables over the viscosity training set, we construct latent variables for unseen intermediate viscosities using the interpolation procedure of Eq.~\eqref{eq:latent_interp}. For example, in Fig.~\ref{fig:general_improvement} we report results at the unseen viscosity value \(\nu=0.28\), for which the IQP generative prediction is reconstructed from \(10^6\) measurement shots. In the top row of Fig.~\ref{fig:general_improvement}, panels (a)--(c) show the numerical ground-truth solutions for the three initial conditions \texttt{dual\_exp}, \texttt{exp}, and \texttt{sin}. The middle row, panels (d)--(f), shows the corresponding predictions obtained with physics-informed latent initialization, while the bottom row, panels (g)--(i), shows the baseline predictions obtained from random latent initialization. The optimization hyperparameters, including learning rate and number of training steps, are kept identical between the baseline and physics-informed settings.

To quantify performance, we use the mean-squared error (MSE) between the reconstructed IQP prediction and the corresponding numerical solution:
\begin{equation}
\mathrm{MSE}
=
\frac{1}{|\Omega|}
\sum_{(x,t)\in\Omega}
\left(
u_{\mathrm{IQP}}(x,t)-u_{\mathrm{GT}}(x,t)
\right)^2,
\end{equation}
where \(\Omega\) denotes the discretized solution domain. In practice, this is computed directly from the ordered prediction and ground-truth arrays. Fig.~\ref{fig:different_low_res} compares the resulting MSE across viscosity values for different surrogate resolutions used to initialize the quantum latent adaptation, namely \(32\times32\), \(16\times16\), and \(12\times12\), together with the random-initialization baseline.

\subsection{Physics-informed initialization improves latent adaptation}
\label{subsec:pinn_init_improves}

Fig.~\ref{fig:general_improvement} presents a qualitative comparison at the unseen viscosity value \(\nu=0.28\). For all three initial conditions, the predictions obtained with physics-informed initialization more closely track the numerical ground truth, whereas the random-initialization baseline shows visibly larger deviations.
To quantify this effect, we evaluate the mean-squared error between the reconstructed IQP prediction and the numerical solution across a range of viscosity values. The results are shown in Fig.~\ref{fig:different_low_res} for the three initial-condition families \texttt{dual\_exp}, \texttt{exp}, and \texttt{sin}. In all cases, the PINN-initialized variants outperform the random-initialization baseline over the tested viscosity range \(\nu \in \{0.13, 0.18, 0.23, 0.28, 0.33, 0.38, 0.43, 0.48, 0.53\}\), showing that the benefit of the proposed initialization is not restricted to a single solution family.

\begin{figure}[t]
    \includegraphics[width=0.48\textwidth]{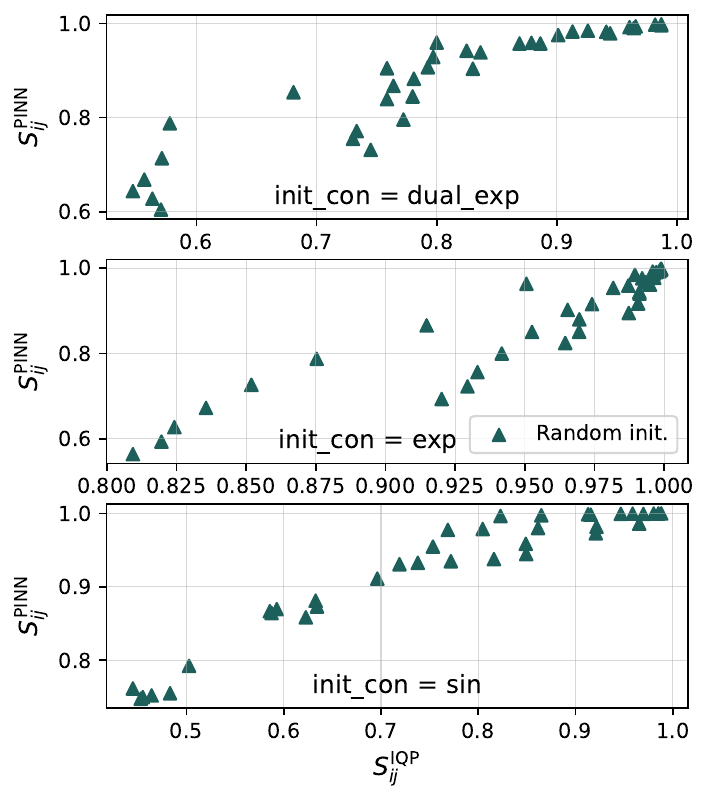}
    \caption{
    Pairwise similarity preservation between the adapted IQP latent space and the classical PINN latent space for the baseline latent-adaptation procedure with random IQP latent initialization. For each viscosity pair \((\nu_i,\nu_j)\), the horizontal coordinate is the IQP latent cosine similarity
    \(S^{\mathrm{IQP}}_{ij}
    =
    \langle \hat{\theta}_{\mathrm{lat}}^{(\nu_i)},
    \hat{\theta}_{\mathrm{lat}}^{(\nu_j)} \rangle\),
    while the vertical coordinate is the corresponding classical latent cosine similarity
    \(S^{\mathrm{PINN}}_{ij}
    =
    \langle \hat{z}^{(\nu_i)},
    \hat{z}^{(\nu_j)} \rangle\).
    The three panels correspond to the Burgers’ equation initial conditions \texttt{dual\_exp}, \texttt{exp}, and \texttt{sin}.}
    \label{fig:similarity_random_init}
\end{figure}

Taken together, these results support the central hypothesis of this work: replacing random latent initialization by a physics-informed latent obtained from a classical surrogate improves the subsequent IQP adaptation process. Empirically, this appears as lower reconstruction error and better qualitative agreement with the target PDE solutions. Methodologically, it suggests that the transferred classical latent places the IQP optimization in a more favorable region of parameter space 
, thereby improving latent adaptation without changing the quantum model architecture, the training objective, or the optimization budget. A more detailed interpretation of
this effect is provided in Sec.~\ref{sec:warm_start_theory}.

\subsection{Cross-resolution transfer from PINN to IQP}
\label{subsec:cross_resolution_transfer}

We next investigate how the quality of the transferred initialization depends on the resolution of the classical surrogate. In Fig.~\ref{fig:different_low_res}, we compare PINN-based initializations obtained from surrogate solution spaces of resolutions \(32\times32\), \(16\times16\), and \(12\times12\). Across all three initial-condition families, even the lowest-resolution surrogate provides a meaningful improvement over the random-initialization baseline, indicating that the transferred prior remains useful even when extracted from substantially coarser classical representations.

At the same time, the dependence on surrogate resolution is not entirely uniform across initial conditions. In the \texttt{exp} case the \(12\times12\) surrogate gives the best MSE among the tested resolutions. This suggests that the usefulness of the transferred latent is not determined solely by surrogate resolution, but may also depend on the structure of the underlying solution family. Overall, the results indicate that cross-resolution transfer from the PINN surrogate to the IQP latent space is robust, and that physics-informed initialization remains beneficial even when the classical representation is learned on a substantially coarser grid.

\subsection{Pairwise similarity preservation between classical and quantum latent spaces}
\label{sec:similarity}

To further examine the PRH-motivated relation in Eq.~\eqref{eq:prh_rel}, we perform an explicit comparison of the pairwise similarity structure induced by the classical surrogate latent variables and the adapted IQP latent variables. Recall that the central hypothesis is not that the two latent coordinates coincide pointwise, but that they preserve a related instance-wise organization of the same underlying PDE solution family. In the present experiment, we make this idea concrete by taking the relation measure \(\operatorname{Rel}(\cdot,\cdot)\) in Eq.~\eqref{eq:prh_rel} to be cosine similarity.

\begin{figure}[t]
    \includegraphics[width=0.48\textwidth]{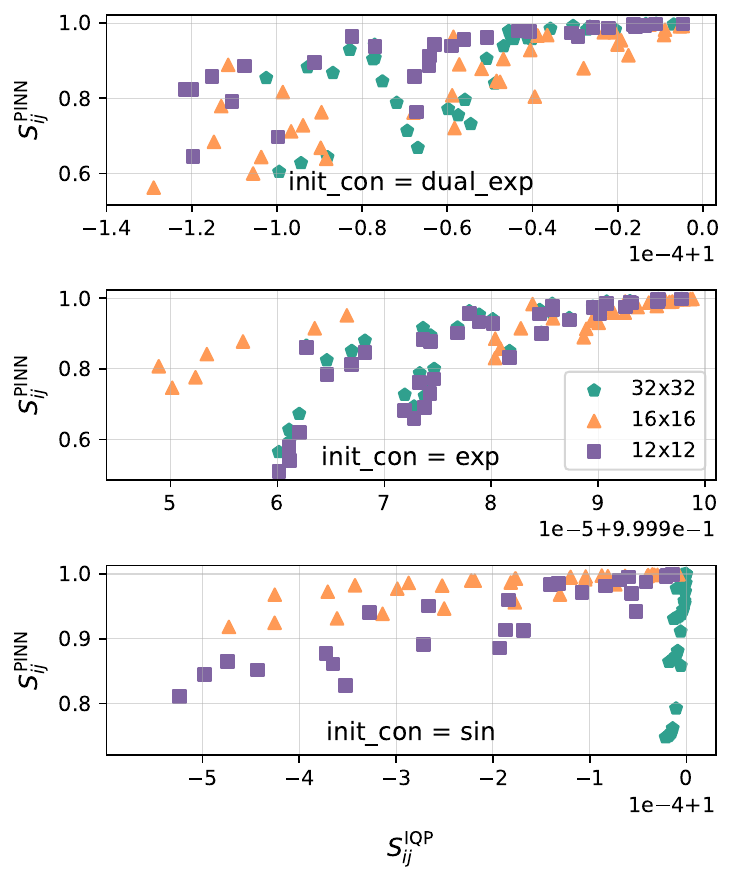}
    \caption{
    Pairwise similarity preservation between the adapted IQP latent space and the classical PINN latent space for physics-informed IQP latent initialization using classical surrogate latents learned at resolutions \(32\times 32\), \(16\times 16\), and \(12\times 12\).
    }
    \label{fig:similarity_pinn_init}
\end{figure}

For each viscosity value \(\nu_i\) in the training set, let \(\theta_{\mathrm{lat}}^{(\nu_i)}\in\mathbb{R}^{d_{\mathrm{lat}}}\) denote the adapted IQP latent variable and let \(z^{(\nu_i)}\in\mathbb{R}^{d_{\mathrm{lat}}}\) denote the corresponding latent variable learned by the classical PINN surrogate. We define the normalized latent vectors
\begin{equation}
\hat{\theta}_{\mathrm{lat}}^{(\nu_i)}
:=
\frac{\theta_{\mathrm{lat}}^{(\nu_i)}}
{\|\theta_{\mathrm{lat}}^{(\nu_i)}\|},
\qquad
\hat{z}^{(\nu_i)}
:=
\frac{z^{(\nu_i)}}{\|z^{(\nu_i)}\|},
\end{equation}
and compute the pairwise cosine similarities
\begin{equation}
S^{\mathrm{IQP}}_{ij}
:=
\left\langle
\hat{\theta}_{\mathrm{lat}}^{(\nu_i)},
\hat{\theta}_{\mathrm{lat}}^{(\nu_j)}
\right\rangle,
\qquad
S^{\mathrm{PINN}}_{ij}
:=
\left\langle
\hat{z}^{(\nu_i)},
\hat{z}^{(\nu_j)}
\right\rangle.
\label{eq:latent_cosine_similarity}
\end{equation}
Here, \(i<j\) ranges over all unordered viscosity pairs, so that each point in the resulting scatter plot corresponds to one pair \((\nu_i,\nu_j)\), with horizontal coordinate \(S^{\mathrm{IQP}}_{ij}\) and vertical coordinate \(S^{\mathrm{PINN}}_{ij}\).

Fig.~\ref{fig:similarity_random_init} first shows the same pairwise-similarity comparison using IQP latent variables obtained from the baseline latent-adaptation procedure with random initialization. Even in this case, the points exhibit a clear positive monotonic trend across all three initial-condition families: viscosity pairs that are more similar in the IQP latent space also tend to be more similar in the classical PINN latent space. This observation is already noteworthy, since it suggests that the learned IQP latent variables are not organized arbitrarily, but instead reflect a relational structure that is broadly compatible with that of the classical surrogate representation.

We then repeat the same analysis for the physics-informed setting in Fig.~\ref{fig:similarity_pinn_init}, where the IQP latent adaptation is initialized by classical surrogate latents obtained at resolutions \(32\times 32\), \(16\times 16\), and \(12\times 12\). Across all three initial-condition families and all tested surrogate resolutions, the scatter plots again show a clear positive monotonic trend. Thus, after physics-informed initialization, viscosity pairs that are more similar in the IQP latent space still tend to be more similar in the classical latent space, indicating that the inter-instance organization is preserved not only qualitatively but also robustly across surrogate resolutions.

A notable difference between the two figures is that, in the physics-informed case, the IQP pairwise cosine similarities are concentrated much closer to \(1\). A natural interpretation is that the transferred PINN initialization has a comparatively large overall magnitude relative to the subsequent latent updates introduced during IQP adaptation. As a result, the adapted IQP latent vectors remain similar in norm and direction, which compresses the cosine similarity values into a narrow high-similarity range. Nevertheless, despite this compression of the horizontal axis, the ordering and organization of the points remain visibly structured, and the positive monotonic trend is still clearly preserved.

Importantly, the observed behavior should be interpreted as approximate relation preservation,
 and supports the weaker and more relevant claim that the relative organization of the viscosity instances is largely preserved across the two representation spaces. This is precisely the type of alignment anticipated in Eq.~\eqref{eq:prh_rel}.

A useful quantitative summary of this effect is given by the Spearman rank correlation \cite{spearman1987proof, forthofer1981rank} (Appendix.~\ref{app:spearman_latent_similarity}) between the sets \(\{S^{\mathrm{IQP}}_{ij}\}_{i<j}\) and \(\{S^{\mathrm{PINN}}_{ij}\}_{i<j}\), since this metric evaluates whether the ordering of pairwise similarities is preserved across the two latent spaces. The resulting values are reported in Table~\ref{tab:spearman_latent_similarity}. Across all initial conditions and surrogate resolutions, the correlations are consistently strong and positive, ranging from \(0.818\) to \(0.969\). These consistently high rank correlations confirm that the relative ordering of pairwise latent similarities is largely preserved between the classical and quantum representations, and further support the interpretation that the classical surrogate latent provides a meaningful structured prior for the IQP latent adaptation procedure.

\begin{table}[t]
\centering
\caption{Spearman rank correlation between the pairwise cosine similarities
\(\{S^{\mathrm{IQP}}_{ij}\}_{i<j}\) and \(\{S^{\mathrm{PINN}}_{ij}\}_{i<j}\),
quantifying the degree to which the ordering of inter-instance similarities is
\vspace{6pt}
preserved between the IQP and classical PINN latent spaces.}
\label{tab:spearman_latent_similarity}
\begin{tabular}{|l|
                S[table-format=1.3]
                S[table-format=1.3]
                S[table-format=1.3]
                S[table-format=1.3]|}
\hline
{Initial condition} & {{32$\times$32}} & {{16$\times$16}} & {{12$\times$12}} & {{Baseline}} \\
\hline
\texttt{dual\_exp} & 0.818 & 0.854 & 0.934 & 0.969 \\
\texttt{exp}       & 0.915 & 0.928 & 0.926 & 0.941 \\
\texttt{sin}       & 0.873 & 0.924 & 0.957 & 0.961 \\
\hline
\end{tabular}
\end{table}

Taken together, the two figures provide an additional empirical perspective on why physics-informed initialization is effective. On the one hand, even the randomly initialized IQP adaptation ultimately learns a latent organization that is positively aligned with the classical surrogate latent space. On the other hand, when the IQP latent is initialized from the classical surrogate, this alignment is retained across multiple surrogate resolutions, while the adapted IQP latents remain concentrated in a narrow high-similarity regime. This suggests that the transferred classical latent does not merely supply a convenient starting vector; rather, it places the subsequent IQP optimization in a latent region whose relational structure is already compatible with that of the target solution family.

To further examine whether this latent-space agreement is an artifact of the experimental design, we include additional ablation studies in Appendix~\ref{app:latent_ablation}. These ablations test possible confounders arising from viscosity distance, a viscosity-only embedding, and the monotonic order used in sequential latent adaptation. Overall, the results show that the observed PINN–IQP latent relation cannot be fully explained by these simple design-induced effects: the alignment persists beyond viscosity-only structure and remains robust to randomized adaptation order for the dual exp and exp cases, while for the sin case the monotonic warm-start order strengthens but does not entirely create the observed correlation. Thus, the appendix results support the interpretation that the measured latent alignment reflects a genuine shared organization between the classical surrogate and IQP latent representations, rather than merely an artifact of the training protocol.

\section{Discussion}
\label{sec:discussion}

\subsection{Distinguishing PDE solution and generative modeling}

It is useful to clarify the different roles played by the PINN surrogate and the IQP model in the present work. A PINN is primarily a physics-informed solver or surrogate model: it represents a solution field through a deterministic function trained to fit data while satisfying the governing differential equation. In contrast, the IQP circuit considered here is a generative model: it defines a probability distribution over discretized solution-domain samples, which are then decoded and aggregated to reconstruct the target field. These two models are therefore not designed to perform the same computational operation, even though they are coupled through a shared physical problem.

From this perspective, the purpose of introducing the IQP model is not to argue that one should use an IQP circuit instead of a PINN to solve Burgers' equation. Indeed, if the only goal were to obtain an accurate solution of this particular PDE, a classical PDE solver or a well-trained PINN would be the more direct tool. Rather, Burgers' equation provides a controlled and interpretable testbed for studying whether a quantum generative model can represent structured scientific data distributions, and whether its latent adaptation can be improved by physics-informed information extracted from a classical surrogate.

The role of the PINN in our method is therefore not to serve as a competing end-to-end generator, but to provide a structured latent representation of the underlying solution family. This latent representation is transferred to initialize the IQP latent variable, after which the IQP model is trained and sampled as a generative model on the higher-resolution solution domain. In this sense, the PINN supplies physics-informed prior information, while the IQP circuit tests whether such information can be used to improve quantum generative learning.

This distinction is important for interpreting the results. The central question addressed by this work is not whether IQP circuits are superior to classical PDE solvers for Burgers' equation. The question is whether classical physics-informed representations can provide useful inductive bias for quantum generative models. The empirical results suggest that they can: transferring the PINN latent representation improves IQP latent adaptation without changing the IQP architecture, training objective, or optimization budget. More broadly, this points to a hybrid methodology in which classical scientific surrogates are used to initialize or guide quantum generative models, especially in settings where the long-term interest is not merely solving a known PDE, but exploring whether quantum generative models can flexibly represent complex scientific data distributions.

\subsection{A warm-start interpretation via effective pre-optimization}
\label{sec:warm_start_theory}

The empirical results of Sec.~\ref{sec:numerics} suggest that the transferred PINN latent does not act as an arbitrary initialization, but rather as a structured warm start
for IQP latent adaptation. A useful way to interpret this intuition is to view
physics-informed initialization as an \emph{effective pre-optimization step} in
the latent optimization landscape.

For a fixed shared IQP core \(\theta_{\mathrm{core}}\) and a given viscosity
\(\nu\), define the latent adaptation objective
\begin{equation}
\mathcal{L}_{\nu}(\theta)
:=
L_{\mathrm{MMD}}\!\bigl(q(\theta_{\mathrm{core}},\theta),\hat p_\nu\bigr),
\label{eq:latent_loss_nu}
\end{equation}
where \(\theta \in \mathbb{R}^{d_{\mathrm{lat}}}\) denotes the instance-dependent
latent variable. Let \(\theta_\nu^* := \theta_{\mathrm{lat}}^{*(\nu)}\) denote a
local minimizer of \(\mathcal{L}_\nu\). To model local optimization dynamics, we
introduce an abstract update operator \(\hat{g}_\nu\), for example a single gradient
step,
\begin{equation}
\hat{g}_\nu(\theta)
=
\theta - \eta \nabla \mathcal{L}_\nu(\theta),
\label{eq:update_operator}
\end{equation}
with stepsize \(\eta > 0\). More generally, \(\hat{g}_\nu\) may be understood as a
local first-order update map associated with the optimizer used in practice.

The central modeling assumption is that the transferred PINN initialization can
be interpreted as an approximate \(m\)-step pre-optimized iterate of the latent
adaptation dynamics:
\begin{equation}
\theta_{\mathrm{lat},0}^{(\nu)}
\approx
\hat{g}_\nu^{\,m}\!\bigl(\theta_{\mathrm{rand},0}^{(\nu)}\bigr),
\qquad m \ge 1,
\label{eq:preopt_assumption}
\end{equation}
where \(\theta_{\mathrm{rand},0}^{(\nu)}\) denotes a random initialization and
\(\hat{g}_\nu^{\,m}\) denotes \(m\) repeated applications of \(\hat{g}_\nu\). In this
interpretation, the transferred classical latent acts not as a generic initial
point, but as though useful optimization progress has already been made before
IQP adaptation begins.

To derive the consequences of this assumption, suppose that \(\hat{g}_\nu\) is locally
contractive around \(\theta_\nu^*\), in the sense that
\begin{equation}
\|\hat{g}_\nu(\theta)-\theta_\nu^*\|
\le
\rho \|\theta-\theta_\nu^*\|,
\qquad 0<\rho<1,
\label{eq:local_contraction}
\end{equation}
for all \(\theta\) in a neighborhood of \(\theta_\nu^*\). By iteration,
Eq.~\eqref{eq:local_contraction} implies
\begin{equation}
\|\hat{g}_\nu^{\,m}(\theta)-\theta_\nu^*\|
\le
\rho^m \|\theta-\theta_\nu^*\|.
\label{eq:iterated_contraction}
\end{equation}

Combining this with the pre-optimization assumption
\eqref{eq:preopt_assumption}, we obtain
\begin{equation}
\|\theta_{\mathrm{lat},0}^{(\nu)}-\theta_\nu^*\|
\approx
\|\hat{g}_\nu^{\,m}(\theta_{\mathrm{rand},0}^{(\nu)})-\theta_\nu^*\|
\le
\rho^m
\|\theta_{\mathrm{rand},0}^{(\nu)}-\theta_\nu^*\|.
\label{eq:distance_reduction}
\end{equation}
Thus, under the effective pre-optimization interpretation, the transferred
initialization is closer to the target minimizer than a random initialization by
a factor controlled by the effective number of pre-optimization steps.

A smaller initialization distance immediately implies a smaller initial loss gap
under standard local regularity conditions. For example, suppose that
\(\mathcal{L}_\nu\) is \(L\)-smooth and satisfies a local Polyak--\L{}ojasiewicz
(PL) inequality with constant \(\mu>0\), namely
\begin{equation}
\frac{1}{2}\|\nabla \mathcal{L}_\nu(\theta)\|^2
\ge
\mu\bigl(\mathcal{L}_\nu(\theta)-\mathcal{L}_\nu(\theta_\nu^*)\bigr),
\label{eq:pl_inequality}
\end{equation}
or, alternatively, a local strong-convexity-type condition. Then distance to the
minimizer controls suboptimality:
\begin{equation}
\mathcal{L}_\nu(\theta)-\mathcal{L}_\nu(\theta_\nu^*)
\le
\frac{L}{2}\|\theta-\theta_\nu^*\|^2.
\label{eq:smoothness_suboptimality}
\end{equation}
Substituting Eq.~\eqref{eq:distance_reduction} into
Eq.~\eqref{eq:smoothness_suboptimality} yields
\begin{equation}
\mathcal{L}_\nu(\theta_{\mathrm{lat},0}^{(\nu)})-\mathcal{L}_\nu(\theta_\nu^*)
\lesssim
\frac{L}{2}\rho^{2m}
\|\theta_{\mathrm{rand},0}^{(\nu)}-\theta_\nu^*\|^2.
\label{eq:initial_gap_bound}
\end{equation}
Therefore, the effective pre-optimization assumption leads directly to a smaller
initial suboptimality for physics-informed initialization.

Finally, under the PL condition, standard first-order convergence results imply exponential decay of the loss gap:
\begin{equation}
\mathcal{L}_\nu(\theta_k)-\mathcal{L}_\nu(\theta_\nu^*)
\le
(1-\eta\mu)^k
\bigl(\mathcal{L}_\nu(\theta_0)-\mathcal{L}_\nu(\theta_\nu^*)\bigr).
\label{eq:exp_convergence}
\end{equation}
Here \(m\) denotes the effective number of optimization steps already encoded in the transferred initialization, whereas \(k\) denotes the subsequent explicit optimization steps carried out during IQP latent adaptation.
Hence, a smaller initial loss gap implies that fewer explicit optimization steps are
required to reach a prescribed accuracy threshold. In this sense, the proposed
physics-informed initialization can be interpreted as placing the IQP latent
adaptation procedure in a more favorable local region of the optimization
landscape.

We emphasize that the above argument should be understood as a local theoretical
interpretation rather than a formal proof that PINN latent transfer is literally
equivalent to running \(m\) IQP optimization steps. Rather, it provides a mechanism consistent with the empirical observations of this work:
the similarity-preservation results suggest that the transferred latent is
structurally aligned with the IQP latent space, and the effective
pre-optimization assumption explains how such alignment can translate into
smaller initialization error, smaller initial loss, and faster convergence.

{
Another natural direction for improving both effectiveness and final performance is to optimize not only the latent initialization, but also the IQP circuit architecture itself. In the present work, the shared IQP core is constructed by including all Pauli-string gates up to locality \(7\) within the \(18\)-qubit system, which provides a broad but not necessarily optimal hypothesis class for the target solution family. A more selective circuit construction, obtained for example through IQP architecture search or structured gate selection, may yield a better inductive-bias match to the underlying problem instance. From this perspective, the success of physics-informed initialization suggests a broader principle: performance depends not only on where optimization begins in parameter space, but also on whether the circuit architecture is well aligned with the correlations and structure of the target data distribution. Identifying such problem-matched IQP architectures, and understanding how architecture choice interacts with latent initialization, therefore remains an important direction for future work.
}

\subsection{Relation to quantum transfer learning and quantum learn-to-learn}

The proposed method is related in spirit to quantum transfer learning \cite{mari2020transfer}, but differs from the standard hybrid-transfer-learning paradigm in an important way. In conventional quantum transfer learning, a pre-trained classical model is often modified or augmented by a trainable quantum component. For example, a classical neural network may be used as a feature extractor, with a variational quantum circuit appended as a final classification layer. In such settings, the classical and quantum components together define the final hybrid model, and the role of the classical network is typically to preprocess high-dimensional data into features suitable for a quantum processor.

In contrast, the present work does not retain the classical PINN surrogate as a preprocessing or postprocessing layer of the final model. The PINN is used only to provide a physics-informed latent representation, which is transferred as the initialization of the IQP latent variable. After this initialization step, the model being optimized and sampled remains the IQP generative model itself. Thus, the final generator is not a hybrid classical--quantum neural network, but an IQP Born machine whose latent parameters have been initialized using information extracted from a classical physics-informed surrogate.

This distinction is important for interpreting the contribution of the method. The goal is not to build a larger hybrid architecture in which a classical network and a quantum circuit jointly perform inference. Rather, the goal is to use a classical surrogate to place a quantum generative model in a more favorable region of its own parameter space before training. In this sense, the transfer occurs at the level of representation and initialization, rather than at the level of model composition.

This perspective also connects naturally to the broader learn-to-learn or meta-learning viewpoint in variational quantum algorithms. In that line of work, the objective is not merely to optimize a circuit for a single instance, but to learn parameters, parameterizations, or update strategies that can be reused across a family of related problems. One example is the meta-VQE framework~\cite{cervera2021meta}, which learns a circuit representation for parameterized Hamiltonians and thereby reduces the optimization burden for subsequent instance-specific tasks. More generally, such approaches suggest that information acquired on one family of problems can be reused to accelerate adaptation on nearby instances.

A particularly relevant direction is parameter transfer across quantum problem instances, especially in QAOA. Previous works have shown that parameters optimized for one instance can often be transferred successfully to other related instances, including larger acceptor problems, when the underlying structure is sufficiently similar~\cite{shaydulin2023parameter,galda2023similarity}. This is conceptually close to the present work in that useful optimization information is reused rather than relearned from scratch. The key difference, however, is that these earlier approaches are primarily quantum-to-quantum: the transferred object is usually a set of variational quantum parameters learned from other quantum instances. By contrast, our method transfers a latent representation learned by a classical physics-informed surrogate into the initialization of a standalone quantum generative model. In this sense, the present approach may be viewed as a classical-to-quantum analogue of the broader learn-to-learn idea, adapted here to scientific generative modeling and cross-resolution latent transfer.

From this perspective, the present method suggests a complementary route to both quantum transfer learning and quantum learn-to-learn. Instead of retaining a classical model as a permanent component of a hybrid architecture, or transferring parameters only between quantum instances, one can use a classical scientific surrogate to provide a structured prior over the latent coordinates of a standalone quantum model. For IQP latent adaptation, this means that the PINN supplies a physics-informed initialization, while the subsequent generative modeling task is still carried out entirely by the IQP circuit distribution itself.
\color{black}

\subsection{Limitations and Future Directions}
\label{sec:limitations}

The present study has several limitations. First, the experiments are restricted to Burgers' equation, which provides a controlled testbed for parameterized scientific data but does not establish universality across more complex PDEs, higher-dimensional systems, or non-PDE datasets.

Second, the present study does not aim to demonstrate quantum advantage or hardware-level quantum utility. Instead, our goal is to study whether physics-informed classical representations can improve the initialization and adaptation of IQP generative models. While classical trainability is an important feature of the IQP learning framework considered here, it remains to be investigated whether the same initialization strategy remains beneficial for larger circuits, finite-shot deployment, noisy devices, or regimes beyond efficient classical simulation.

Third, the mechanism underlying successful latent transfer is not yet fully characterized. The pairwise similarity analysis and ablation studies support the view that the PINN and IQP latent spaces share a compatible inter-instance organization, but this remains an empirical observation rather than a complete theory. The effectiveness of the method may depend on the surrogate quality, latent dimension, circuit architecture, data encoding, parameter ordering, and the structural match between the target distribution and the IQP model family. Developing principled diagnostics for when classical-to-quantum latent transfer should succeed therefore remains an open problem for future investigation.

\section{Conclusion}
\label{sec:conclusion}

In this work, we introduced a physics-informed latent initialization scheme for generative IQP circuit learning. The central idea is to use a classical physics-informed surrogate as a source of structured latent information for initializing the quantum generative model. Concretely, we trained a latent-conditioned PINN surrogate on lower-resolution Burgers' equation solution data, extracted its learned latent representation for a reference viscosity, and transferred this representation to initialize the latent variables of an IQP Born machine. The IQP model was then trained and adapted on the higher-resolution solution domain using the standard latent-adaptation protocol.

Our numerical experiments show that this physics-informed initialization consistently improves IQP latent adaptation relative to random initialization. Across three Burgers' equation initial-condition families and multiple unseen viscosity values, the PINN-initialized IQP models achieve lower reconstruction error and better qualitative agreement with the numerical solutions. The improvement persists even when the classical surrogate is trained on substantially coarser grids, indicating that the transferred latent representation provides a robust prior. Pairwise similarity analysis further shows that the classical PINN latent space and the adapted IQP latent space preserve a compatible inter-instance organization, supporting the interpretation that the transferred latent places the IQP optimization in a more favorable and physically meaningful region of parameter space.
{Although the present experiments focus on viscosity in Burgers’ equation, the proposed method is not specific to this particular parameter or PDE, and in principle can be extended to other parameterized physical systems for which a classical surrogate can provide structured latent representations.}

We further compared the proposed IQP-PINN latent initialization scheme with conditional DCGAN baselines in Appendix~\ref{sec:classical_baselines}. These results show that the relative performance depends strongly on the solution family and on the inductive-bias match between model class and data distribution. The IQP-PINN model performs strongly for the \texttt{dual\_exp} case and remains competitive for \texttt{sin}, while the cDCGAN baselines outperform it for \texttt{exp}. This comparison should therefore be interpreted as a reference baseline rather than evidence for a general separation between quantum and classical generative models. More broadly, it reinforces that identifying when a given quantum generative architecture is well matched to a scientific dataset remains an important open problem.

These results suggest a broader perspective on how latent representations may be used across classical and quantum learning systems. Instead of viewing classical and quantum models as isolated alternatives, one can use classical models to provide structured representations, warm starts, or inductive biases for quantum models. In future settings where a quantum model cannot be efficiently simulated classically, the best available classical model may still provide valuable information for initializing or organizing the quantum parameter space. This offers a possible route for connecting state-of-the-art classical modeling with quantum generative learning: classical models can guide the quantum model toward meaningful regions of parameter space, while quantum models may then be used to investigate regimes that are increasingly difficult to reach with classical methods alone. Understanding this interface between classical representations and quantum generative models is an important direction for future work.

\begin{acknowledgments}
We are grateful to Stephen Clark, Konstantinos Meichanetzidis, and Daniel Mills for helpful discussions.
\end{acknowledgments}

\bibliography{refs}

\clearpage
\onecolumngrid
\appendix

\section{Geometric motivation for latent adaptation}
\label{app:latent_adaptation_motivation}

In this appendix, we provide a simple geometric motivation for the latent-adaptation assumption used in the main text. The purpose is to clarify why adapting only a low-dimensional latent block is natural when the target objects vary smoothly with an external physical parameter.

Let \(\nu \in \mathcal{V} \subset \mathbb{R}\) denote a physical parameter, such as the viscosity in Burgers' equation, and let \(p_\nu\) denote the target distribution over discretized solution-domain samples associated with that parameter. In the present work, \(p_\nu\) is obtained by mapping the continuous solution field into the bitstring domain through the float-to-bit encoding described in Sec.~\ref{subsec:latent_adaptation}. The basic latent-adaptation assumption is that the family
\begin{equation}
    \mathcal{M}_{\rm data}
    :=
    \{p_\nu : \nu \in \mathcal{V}\}
\end{equation}
forms a low-dimensional and sufficiently smooth manifold of distributions.

The IQP latent-adaptation model represents this family by fixing a shared circuit core \(\theta_{\rm core}\) and varying only an instance-dependent latent block \(\theta_{\rm lat}\). For a fixed core, the IQP model therefore defines a latent distribution map
\begin{equation}
    F_{\theta_{\rm core}} :
    \mathbb{R}^{d_{\rm lat}}
    \rightarrow
    \mathcal{P}(\{0,1\}^n),
    \qquad
    \theta_{\rm lat}
    \mapsto
    q_{\theta_{\rm core},\theta_{\rm lat}},
\end{equation}
where \(q_{\theta_{\rm core},\theta_{\rm lat}}\) denotes the bitstring distribution generated by the IQP circuit. Latent adaptation assumes that, over the parameter range of interest, there exists a latent trajectory
\begin{equation}
    \gamma :
    \mathcal{V}
    \rightarrow
    \mathbb{R}^{d_{\rm lat}},
    \qquad
    \nu
    \mapsto
    \theta_{\rm lat}^{(\nu)},
\end{equation}
such that
\begin{equation}
    q_{\theta_{\rm core},\gamma(\nu)}
    \approx
    p_\nu .
\end{equation}
In other words, the shared core captures structure common to the family, while the latent block parameterizes the variation across different physical instances.

This assumption implies that nearby physical parameters should require only small latent changes. Suppose that the latent trajectory \(\gamma\) is differentiable and has bounded derivative,
\begin{equation}
    \left\|
    \frac{d\gamma}{d\nu}
    \right\|
    \leq
    L_\gamma .
\end{equation}
Then, for two nearby parameter values \(\nu\) and \(\nu+\Delta\nu\), we have
\begin{align}
    \left\|
    \theta_{\rm lat}^{(\nu+\Delta\nu)}
    -
    \theta_{\rm lat}^{(\nu)}
    \right\|
    &=
    \left\|
    \gamma(\nu+\Delta\nu)
    -
    \gamma(\nu)
    \right\|  \nonumber \\
    &\leq
    L_\gamma |\Delta\nu| .
\end{align}
Thus, under a smooth latent parameterization, small changes in the physical parameter correspond to small changes in the latent variables. This provides a natural justification for sequential latent adaptation: after fitting one instance, the optimized latent variable provides a useful warm start for the next nearby instance.

The same intuition can be expressed directly at the level of distributions. Let \(D(\cdot,\cdot)\) be a metric or discrepancy between probability distributions, such as MMD. Suppose that the target family is Lipschitz continuous (with constant $L_p$) in this discrepancy,
\begin{equation}
    D(p_{\nu+\Delta\nu},p_\nu)
    \leq
    L_p |\Delta\nu| ,
\end{equation}
and that the fixed-core IQP latent map is locally Lipschitz,
\begin{equation}
    D(
    q_{\theta_{\rm core},\theta_{\rm lat}+\Delta\theta},
    q_{\theta_{\rm core},\theta_{\rm lat}}
    )
    \leq
    L_q \|\Delta\theta\| .
\end{equation}
If the fixed-core IQP family is locally expressive enough to follow the data manifold, then tracking the change from \(p_\nu\) to \(p_{\nu+\Delta\nu}\) should require a latent displacement whose size scales with the parameter displacement,
\begin{equation}
    \|
    \Delta\theta_{\rm lat}
    \|
    =
    O(|\Delta\nu|).
\end{equation}
This geometric picture explains why it can be effective to freeze the shared core and adapt only a low-dimensional latent block across nearby physical instances.

This perspective also clarifies the role of physics-informed initialization. A classical surrogate, such as a latent-conditioned PINN, learns its own latent representation of the same parameterized solution family. If the classical and quantum latent representations preserve a compatible relational organization, then the classical latent can place the IQP latent variable near the appropriate region of the quantum latent manifold. Subsequent IQP adaptation then only needs to refine this structured initialization within the quantum model family, rather than discovering the latent organization from an uninformed random starting point.

\section{Spearman rank correlation for pairwise latent similarities}
\label{app:spearman_latent_similarity}

In Sec.~\ref{sec:similarity}, we compare the pairwise cosine similarities \(\{S^{\mathrm{IQP}}_{ij}\}_{i<j}\) and \(\{S^{\mathrm{PINN}}_{ij}\}_{i<j}\) in order to assess whether the inter-instance relational structure is preserved between the IQP latent space and the classical PINN latent space. As a quantitative summary of this effect, we use the Spearman rank correlation \cite{spearman1987proof, forthofer1981rank}.

Given two collections of scalar values \[ \mathcal{X}=\{X_k\}_{k=1}^{M}, \qquad \mathcal{Y}=\{Y_k\}_{k=1}^{M}, \] the Spearman rank correlation measures the agreement between their rank orderings rather than their raw numerical values. In our setting, these two collections are \[ \mathcal{X}=\{S^{\mathrm{IQP}}_{ij}\}_{i<j}, \qquad \mathcal{Y}=\{S^{\mathrm{PINN}}_{ij}\}_{i<j}, \] where each entry corresponds to one unordered viscosity pair \((\nu_i,\nu_j)\).
Let \(R(X_k)\) and \(R(Y_k)\) denote the ranks of \(X_k\) and \(Y_k)\), respectively, among their corresponding collections. The Spearman rank correlation is then defined as the Pearson correlation coefficient between the ranks:
\begin{equation} \rho_{\mathrm{S}} = \frac{ \sum_{k=1}^{M} \bigl(R(X_k)-\overline{R_X}\bigr) \bigl(R(Y_k)-\overline{R_Y}\bigr) }{ \sqrt{ \sum_{k=1}^{M}\bigl(R(X_k)-\overline{R_X}\bigr)^2 } \sqrt{ \sum_{k=1}^{M}\bigl(R(Y_k)-\overline{R_Y}\bigr)^2 } },
\label{eq:spearman_def}
\end{equation}

where \(\overline{R_X}\) and \(\overline{R_Y}\) are the mean ranks of the two collections.
The key reason for using \(\rho_{\mathrm{S}}\) here is that our goal is not to test exact equality or exact linear proportionality between \(S^{\mathrm{IQP}}_{ij}\) and \(S^{\mathrm{PINN}}_{ij}\). Rather, we wish to evaluate whether the \emph{ordering} of pairwise similarities is preserved across the two latent spaces. A high positive value of \(\rho_{\mathrm{S}}\) therefore indicates that viscosity pairs that are more similar in the IQP latent space also tend to be more similar in the classical PINN latent space.
The interpretation of \(\rho_{\mathrm{S}}\) is standard: 
\begin{itemize}
    \item $\rho_{\mathrm{S}}=1 \quad \text{corresponds to perfect rank agreement,}$
    \item $\rho_{\mathrm{S}}=0 \quad \text{corresponds to no monotonic association,}$
    \item $\rho_{\mathrm{S}}=-1 \quad \text{corresponds to perfectly reversed rank ordering.}$
\end{itemize}
Thus, in the present context, consistently large positive values of \(\rho_{\mathrm{S}}\) support the claim that the pairwise relational structure of the viscosity instances is largely preserved between the classical and quantum latent representations.
The Spearman rank correlation values reported in Table~\ref{tab:spearman_latent_similarity} are all strongly positive, providing quantitative support for the monotonic trends observed in Figs.~4 and~5.

\section{Ablation Studies of Latent Relation}
\label{app:latent_ablation}

In the main text, we observe that the pairwise similarity structure of the learned IQP latent variables is strongly aligned with that of the classical PINN latent representations. Since both latent trajectories are indexed by the same physical parameter, viscosity, it is important to check whether this agreement reflects a genuine shared latent organization or is instead induced by experimental design choices. In particular, three possible confounders must be considered: nearby viscosities naturally produce similar PDE solutions, a trivial one-dimensional viscosity embedding may already explain much of the similarity structure, and sequential latent adaptation in monotonic viscosity order may itself impose an ordered latent trajectory.

We therefore perform three ablation studies. Each ablation compares pairwise cosine similarities between PINN latents and IQP latent variables as in main text.
The purpose of these tests is not to prove that viscosity plays no role, since it clearly parameterizes the solution family, but rather to determine whether the observed PINN-IQP relation contains signal beyond simple viscosity ordering or training-order artifacts.

\subsection{Viscosity-Distance Control}

First, we test whether the PINN-IQP latent alignment remains after explicitly controlling for viscosity distance. For every pair of viscosities, we compute
\begin{equation}
D_{ij} = |\nu_i - \nu_j|,
\end{equation}
together with $S^{\mathrm{PINN}}_{ij}$ and $S^{\mathrm{IQP}}_{ij}$. 
We then compute the raw Spearman correlation
\begin{equation}
\rho\!\left(S^{\mathrm{PINN}}, S^{\mathrm{IQP}}\right),
\end{equation}
as well as the partial Spearman correlation obtained by rank-transforming all variables and regressing out the effect of \(D_{ij}\). Here, the rank transform is applied to the full vector of pairwise observations. For example, the set \(\{D_{ij}: i<j\}\) contains all pairwise viscosity distances. Then \(\operatorname{rank}(D_{ij})\) denotes the position of \(D_{ij}\) within this set after sorting the distances from smallest to largest. Tied values are assigned their average rank, as in the standard Spearman rank correlation procedure. Thus, \(\operatorname{rank}(D_{ij})\) ranks that pairwise distance relative to all other viscosity pairs.
Specifically, for the PINN similarities we set
\[
x_{ij} = \operatorname{rank}(D_{ij}),
\qquad
y^{\mathrm{PINN}}_{ij}
=
\operatorname{rank}(S^{\mathrm{PINN}}_{ij}),
\]
and fit
\[
y^{\mathrm{PINN}}_{ij}
=
\beta^{\mathrm{PINN}}_0
+
\beta^{\mathrm{PINN}}_1 x_{ij}
+
\epsilon^{\mathrm{PINN}}_{ij}.
\]
The residual is then
\[
r^{\mathrm{PINN}}_{ij}
=
y^{\mathrm{PINN}}_{ij}
-
\left(
\widehat{\beta}^{\mathrm{PINN}}_0
+
\widehat{\beta}^{\mathrm{PINN}}_1 x_{ij}
\right).
\]
We compute $r^{\mathrm{IQP}}_{ij}$ analogously by replacing
$S^{\mathrm{PINN}}_{ij}$ with $S^{\mathrm{IQP}}_{ij}$. The reported partial Spearman correlation is the Pearson correlation between these two residual vectors,
\[
\rho_{\mathrm{partial}}
=
\operatorname{corr}
\left(
r^{\mathrm{PINN}},
r^{\mathrm{IQP}}
\right).
\]
This measures whether the PINN and IQP latent similarity structures remain aligned after subtracting the component that is linearly explained, in rank space, by viscosity distance.

\begin{figure}[t]
    \centering
    \includegraphics[width=\linewidth]{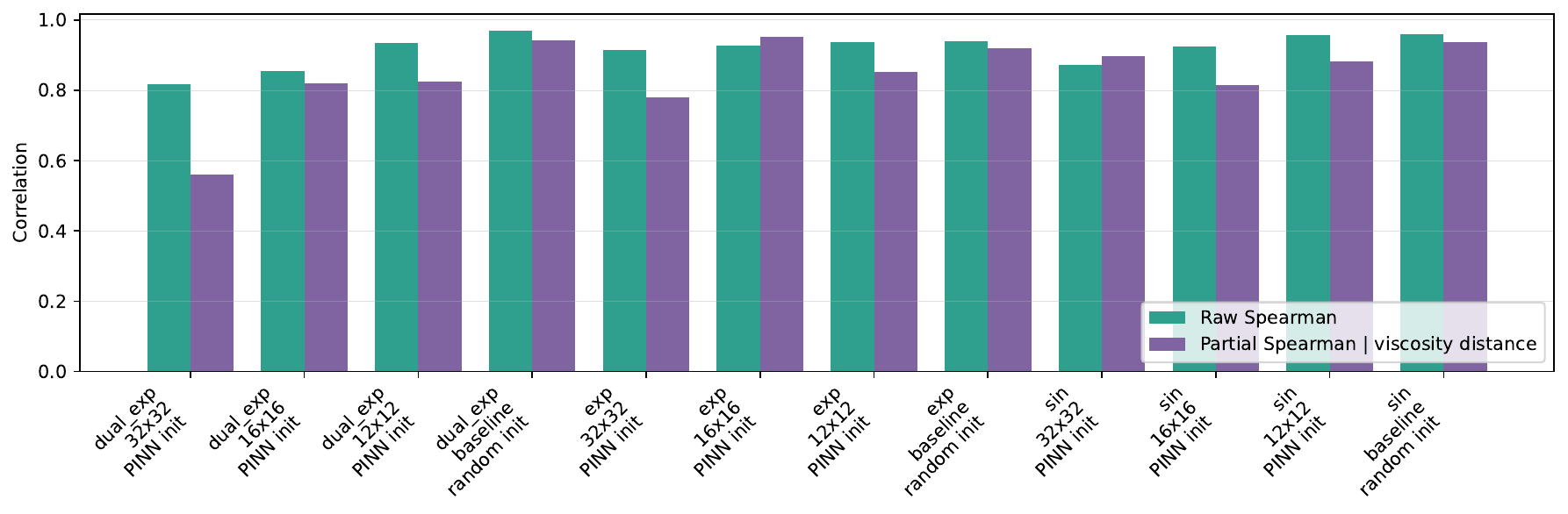}
    \caption{
    Viscosity-distance control. Raw Spearman correlations between pairwise PINN and IQP latent similarities are compared with partial Spearman correlations after controlling for pairwise viscosity distance. The controlled correlations remain strongly positive, indicating that the latent alignment is not solely a consequence of nearby viscosities being similar.
    }
    \label{fig:latent_ablation_distance_control}
\end{figure}

The results are shown in Fig.~\ref{fig:latent_ablation_distance_control}. The partial correlations remain strongly positive after removing the monotonic effect of viscosity distance. This indicates that the observed PINN-IQP latent relation is not explained solely by the fact that nearby viscosities correspond to similar Burgers solutions. Instead, the two latent spaces preserve additional shared relational structure.

\subsection{Trivial Viscosity-Only Baseline}

Second, we compare the learned latent similarities against a trivial one-dimensional viscosity embedding. In this baseline, each viscosity is represented only by itself,
\begin{equation}
z_{\mathrm{visc}}(\nu_i) = [\nu_i],
\end{equation}
and pairwise similarity is defined as a monotonic function of viscosity distance. We use
\begin{equation}
S^{\mathrm{visc}}_{ij} = -|\nu_i - \nu_j|,
\end{equation}
although other monotonic choices, such as $\exp(-\gamma |\nu_i-\nu_j|)$ or $(1+|\nu_i-\nu_j|)^{-1}$, give the same Spearman ranking.

\begin{figure}[t]
    \centering
    \includegraphics[width=\linewidth]{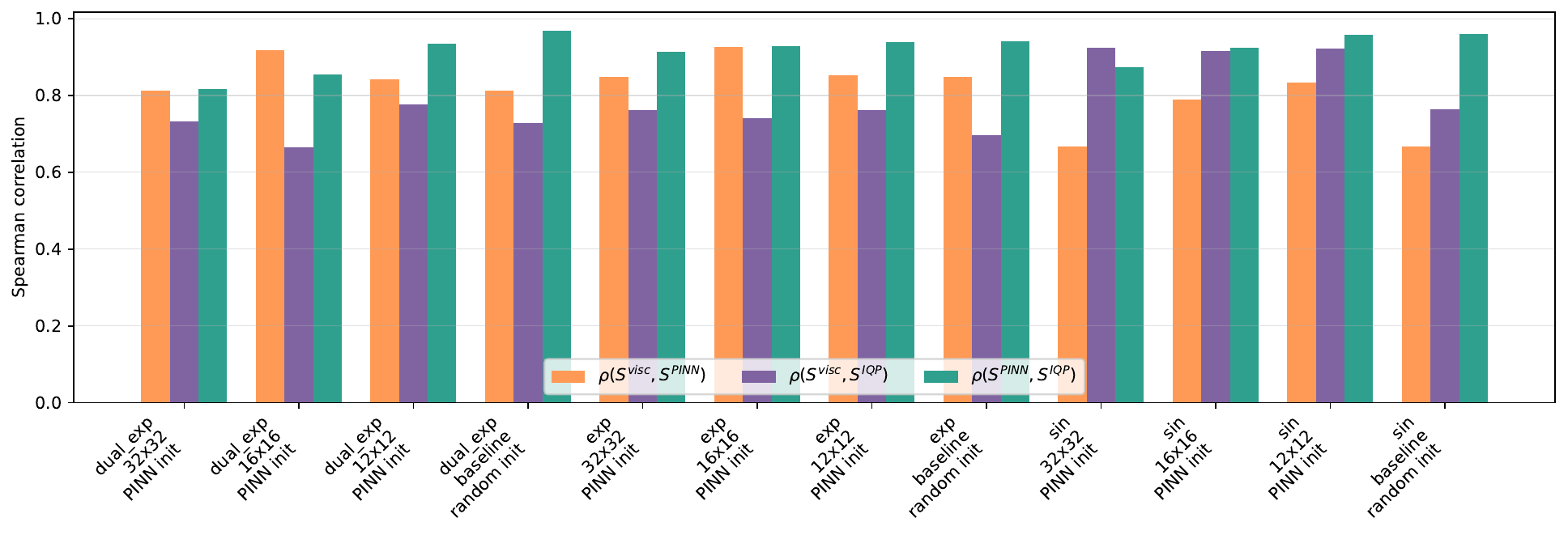}
    \caption{
    Trivial viscosity-only baseline. Pairwise similarities induced by a one-dimensional viscosity embedding are compared with pairwise PINN and IQP latent similarities. The viscosity-only baseline captures part of the structure, but does not fully account for the learned PINN-IQP latent relation.
    }
    \label{fig:latent_ablation_viscosity_baseline}
\end{figure}

We then compare
\begin{equation}
\rho(S^{\mathrm{visc}}, S^{\mathrm{PINN}}),
\quad
\rho(S^{\mathrm{visc}}, S^{\mathrm{IQP}}),
\quad
\rho(S^{\mathrm{PINN}}, S^{\mathrm{IQP}}).
\end{equation}
The results are shown in Fig.~\ref{fig:latent_ablation_viscosity_baseline}. As expected, viscosity alone explains a substantial part of the similarity structure, since viscosity is the physical parameter controlling the solution family. However, the learned PINN-IQP correlations are not uniformly reducible to this one-dimensional baseline. In particular, the relative behavior differs across initial conditions, suggesting that the learned latent spaces encode more than a trivial ordering by viscosity.

The random-initialization baselines reveal an additional feature of this
comparison. In all three initial-condition families, the direct latent
correlation
\(\rho(S^{\mathrm{PINN}},S^{\mathrm{IQP}})\)
for the random-init baseline is substantially larger than both
\(\rho(S^{\mathrm{visc}},S^{\mathrm{PINN}})\) and
\(\rho(S^{\mathrm{visc}},S^{\mathrm{IQP}})\). Thus, although the random-init
IQP latents can exhibit a strong pairwise agreement with the PINN latent
geometry, this agreement is not captured as directly by the trivial
one-dimensional viscosity baseline.

By contrast, in the PINN-initialized runs, the IQP-viscosity correlation
\(\rho(S^{\mathrm{visc}},S^{\mathrm{IQP}})\) becomes more comparable to the
PINN-viscosity correlation
\(\rho(S^{\mathrm{visc}},S^{\mathrm{PINN}})\). This suggests that transferring
the PINN latent representation does not merely increase the raw PINN-IQP
similarity correlation; rather, it anchors the IQP latent trajectory more
closely to the physically meaningful viscosity ordering already encoded by the
classical surrogate. In this sense, the PINN initialization appears to inject
viscosity-structured information into the IQP latent variables.
Therefore, the viscosity-only baseline indicates that viscosity ordering is an
important component of the transferred latent structure. The PINN initialization
appears to make the IQP latent geometry more consistent with this physical
one-dimensional organization, whereas the random-init latents can show high
PINN-IQP similarity without being as directly explained by the viscosity-only
baseline.

\subsection{Randomized Adaptation Order}

Finally, we test whether the latent relation is induced by the monotonic warm-start procedure used during sequential IQP latent adaptation. In the original experiment, viscosities are adapted in increasing order,
\begin{equation}
\nu_0,\nu_1,\ldots,\nu_K.
\end{equation}
This could impose a smooth trajectory in IQP latent space even if the learned relation were partly an artifact of training order.

To test this, we repeat the IQP latent-adaptation experiment with randomized viscosity orders. For each random seed, we shuffle the viscosity sequence, train the IQP latents sequentially in that shuffled order, and then reorder the learned latents back into the standard viscosity order before computing pairwise similarities. We use five shuffled seeds for each initial condition. We then compare the original monotonic-order correlation against the mean and standard deviation of the shuffled-order correlations:
\begin{equation}
\rho_{\mathrm{orig}}
=
\rho(S^{\mathrm{PINN}}, S^{\mathrm{IQP}}_{\mathrm{orig}}),
\end{equation}
and
\begin{equation}
\rho_{\mathrm{shuf}}
=
\rho(S^{\mathrm{PINN}}, S^{\mathrm{IQP}}_{\mathrm{shuf}}).
\end{equation}

\begin{figure}[t]
    \centering
    \includegraphics[width=0.5\linewidth]{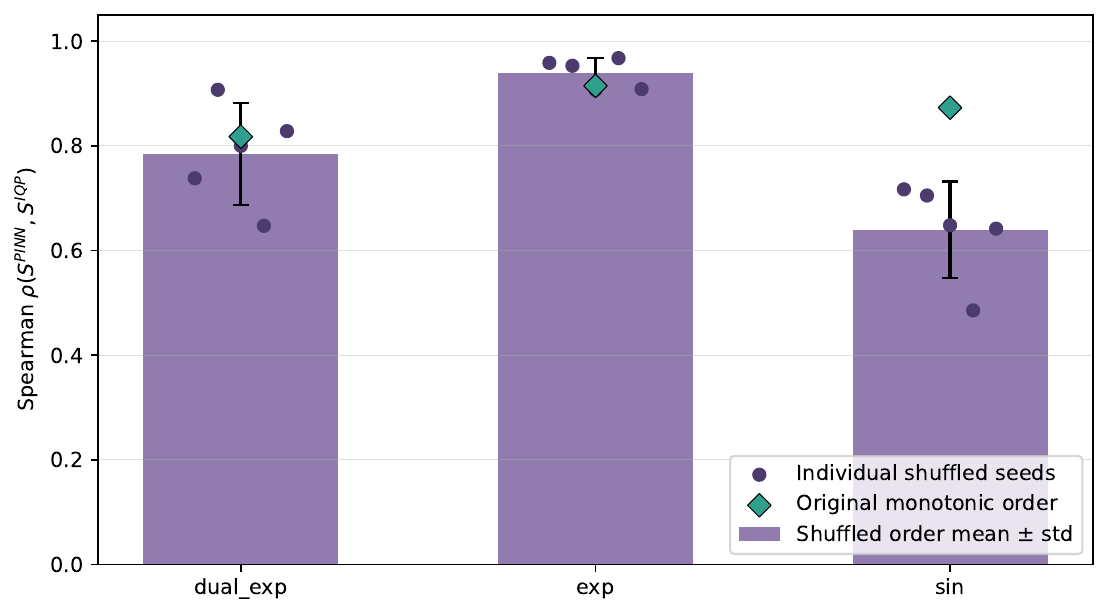}
    \caption{
    Randomized adaptation-order ablation. The original monotonic-order PINN-IQP latent correlation is compared with shuffled-order runs over five random seeds. The alignment is robust for \texttt{dual\_exp} and \texttt{exp}; for \texttt{sin}, monotonic warm-start training strengthens the alignment, but positive correlation remains under shuffled adaptation.
    }
    \label{fig:latent_ablation_order}
\end{figure}

The results are shown in Fig.~\ref{fig:latent_ablation_order}. For the \texttt{dual\_exp} and \texttt{exp} initial conditions, the shuffled-order correlations remain close to the original monotonic-order correlations, indicating that the latent alignment is robust to adaptation order. For the \texttt{sin} initial condition, randomizing the order reduces the mean correlation, suggesting that monotonic warm-start training contributes to the observed alignment in that case. Nevertheless, the shuffled-order correlations remain positive, indicating that the latent relation is weakened but not eliminated.

Overall, these ablations support the interpretation that the observed PINN-IQP latent relation is a real signal rather than a simple artifact of viscosity distance, viscosity-only ordering, or monotonic training order. At the same time, the randomized-order ablation shows that the strength of the effect can depend on the structure of the underlying solution family.

\section{Classical Baselines}
\label{sec:classical_baselines}

To contextualize the performance of the proposed physics-informed IQP latent initialization scheme, we additionally compare against a selected family of classical generative baselines. The purpose of this comparison is not to exhaust the space of possible classical generative models, but rather to provide a reference point for a widely used convolutional generative architecture trained directly on the same Burgers' equation solution families.

\begin{figure*}[t]
    \centering
    \includegraphics[width=\textwidth]{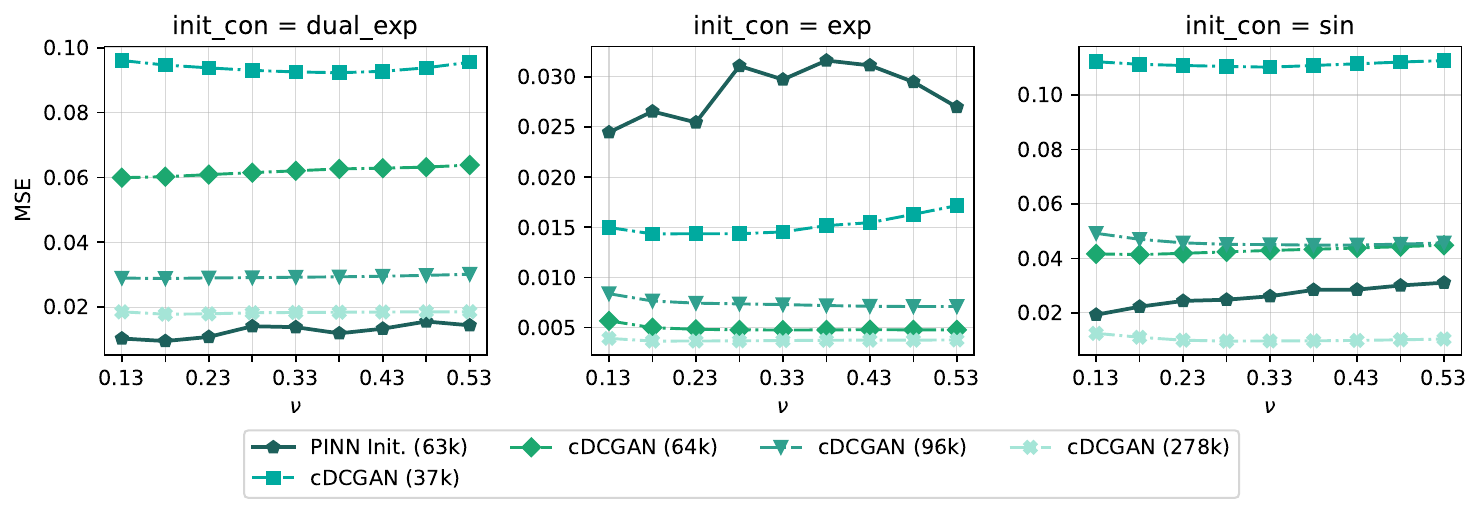}
    \caption{
    Classical cDCGAN baselines for Burgers' equation solution generation. 
    We compare the IQP model initialized from a $12\times 12$ PINN surrogate, with approximately $63$k trainable parameters, against conditional DCGAN baselines with approximately $37$k, $64$k, $96$k, and $278$k trainable parameters. 
    Each model is evaluated on unseen viscosity values $\nu \in \{0.13,0.18,0.23,0.28,0.33,0.38,0.43,0.48,0.53\}$ using the MSE between the generated and numerical Burgers' equation solution fields. 
    The IQP-PINN model achieves the best performance for the \texttt{dual\_exp} initial condition and remains competitive for \texttt{sin}, while the cDCGAN baselines outperform it for the \texttt{exp} initial condition. 
    These results indicate that the relative performance depends on the inductive-bias match between model class and solution family, and that the present comparison should be understood as a reference baseline rather than an exhaustive search over classical generative architectures.
    }
    \label{fig:classical_baselines_dcgan}
\end{figure*}

\subsection{Conditional DCGAN baseline}

We use a conditional deep convolutional generative adversarial network (cDCGAN) \cite{radford2016} as the classical generative baseline. In contrast to the IQP model, which represents samples in the binarized $(x,t,u)$ domain and reconstructs the solution field through sampling and inverse decoding, the cDCGAN treats each Burgers' equation solution as a three-channel image. Specifically, each training example is represented as
\begin{equation}
    I_\nu = \big[X, T, U_\nu(X,T)\big] \in \mathbb{R}^{3 \times 64 \times 64},
\end{equation}
where the three channels correspond to the spatial grid, temporal grid, and solution value, respectively. The viscosity $\nu$ is used as a conditioning variable. Both the solution image and the viscosity label are normalized to the interval $[-1,1]$ before training.

The generator takes as input a random latent vector $z \in \mathbb{R}^{d_z}$ together with the normalized viscosity $\nu$, and maps this conditional latent input to a full $3 \times 64 \times 64$ solution-domain image:
\begin{equation}
    G_{\phi}(z,\nu) \mapsto \widehat{I}_\nu .
\end{equation}
The discriminator receives either a real or generated image, concatenated with a spatially broadcast viscosity-conditioning channel, and predicts whether the image is real or generated. The cDCGAN is trained with the standard adversarial binary cross-entropy objective using Adam optimizers for both generator and discriminator. We use learning rates $2\times 10^{-4}$, Adam parameters $(\beta_1,\beta_2)=(0.5,0.999)$, batch size $10$, and label smoothing of $0.1$ for the real labels. Each model is trained for $5000$ epochs on the ten training viscosities
\begin{equation}
\nu \in \{0.06,0.11,0.16,0.21,0.26,0.31,0.36,0.41,0.46,0.51\}.
\end{equation}
To examine the dependence on classical model capacity, we vary the generator and discriminator widths, resulting in cDCGAN baselines with approximately $37$k, $64$k, $96$k, and $278$k trainable parameters. We compare these baselines against the IQP model initialized by the $12\times 12$ PINN surrogate (same as in Fig.~\ref{fig:different_low_res}), which has approximately $63$k trainable parameters.

For evaluation, the trained cDCGAN is conditioned on the same unseen viscosity values used in the IQP interpolation experiment,
\begin{equation}
\nu \in \{0.13,0.18,0.23,0.28,0.33,0.38,0.43,0.48,0.53\}.
\end{equation}
The generated solution channel is extracted from $\widehat{I}_\nu$ and compared with the corresponding numerical Burgers' equation solution using the same mean-squared error metric as in Sec.~\ref{subsec:exp_setup}.

\subsection{Comparison with IQP-PINN latent adaptation}

The results are shown in Fig.~\ref{fig:classical_baselines_dcgan}. The comparison shows that the relative performance of the IQP-PINN model and the cDCGAN baselines depends strongly on the initial-condition family.

For the \texttt{dual\_exp} initial condition, the IQP model with $12\times 12$ PINN initialization achieves the lowest MSE across the tested viscosity range. Among the cDCGAN baselines, increasing the model size improves performance: the smallest $37$k-parameter cDCGAN has the largest error, while the $278$k-parameter cDCGAN is substantially better. Nevertheless, even the largest cDCGAN baseline remains above the IQP-PINN curve in this case.

For the \texttt{sin} initial condition, the IQP-PINN model again performs competitively and remains better than the smaller and intermediate cDCGAN baselines. The largest cDCGAN baseline achieves the lowest MSE among the tested models, indicating that sufficient classical capacity can be highly effective for this solution family. However, among models of comparable parameter scale, the $63$k-parameter IQP-PINN model performs favorably relative to the $37$k, $64$k, and $96$k cDCGAN baselines.

The \texttt{exp} initial condition exhibits the opposite trend. In this case, the cDCGAN baselines, especially the $64$k-, $96$k-, and $278$k-parameter models, outperform the IQP-PINN model across the tested viscosity values. This indicates that the advantage of physics-informed IQP latent adaptation is not uniform across all solution families. Rather, the effectiveness of a given generative model depends on the structural match between the model class, the training protocol, and the target data distribution.

These results should therefore be interpreted as a model- and dataset-dependent comparison rather than evidence for a general separation between quantum and classical generative models. The cDCGAN baseline demonstrates that classical convolutional generators can be highly competitive, and in some cases superior, on the same Burgers' equation interpolation task. Conversely, the IQP-PINN model performs strongly on \texttt{dual\_exp} and remains competitive on \texttt{sin}, despite using a compact latent-adaptation mechanism and a physics-informed initialization extracted from a substantially lower-resolution classical surrogate.

\subsection{Scope of the classical comparison}

We emphasize that these experiments do not rule out the existence of stronger classical baselines. Modern classical generative modeling includes many architectures beyond DCGANs, including U-Nets, diffusion models, transformer-based generators, neural operators, and autoregressive models. A comprehensive search over classical architectures, quantum architectures, optimization procedures, and data representations is beyond the scope of this work.

This limitation is conceptually important. The broader question of when a quantum generative model is well matched to a given dataset, and when a classical model provides a better inductive bias, remains open. Previous work has proposed dataset-level diagnostics such as quantum-correlation-likeness and classical correlation-complexity indicators to study this type of inductive-bias matching \cite{liu2026toward}. However, establishing such a full model--dataset selection theory is not the primary goal of the present work.

Instead, the goal of this work is narrower: we study whether a physics-informed classical latent representation can improve IQP latent adaptation. From this perspective, the classical baselines in Fig.~\ref{fig:classical_baselines_dcgan} serve as reference points rather than exhaustive competitors. The results support a balanced conclusion: physics-informed initialization can make IQP latent adaptation highly competitive for some structured PDE solution families, but we do not claim that the resulting IQP model outperforms all possible classical generative approaches on this task.

\section{Reproducibility Summary}
\label{app:reproducibility}

This appendix summarizes the main numerical settings used throughout the paper. The key data-generation parameters, model configurations, optimization settings, ablation protocols, and evaluation procedures are collected here for reproducibility and interpretability.

Tables~\ref{tab:repro_burgers_data}--\ref{tab:repro_iqp} specify the Burgers' equation dataset, bitstring encoding, PINN surrogate, latent-transfer procedure, IQP circuit, and optimization hyperparameters. Table~\ref{tab:repro_main_runs} gives the main IQP experiment matrix, including the random-initialization baseline and the three PINN-initialized variants for each initial-condition family. Tables~\ref{tab:repro_latent_ablation} and~\ref{tab:repro_order_ablation} summarize the latent-relation ablations, including the randomized adaptation-order protocol. Table~\ref{tab:repro_dcgan} reports the conditional DCGAN baseline settings, and Table~\ref{tab:repro_seeds} records the random-seed usage and numerical reproducibility caveats.

\begin{table}[h]
\centering
\caption{Burgers' equation dataset and bitstring representation.}
\label{tab:repro_burgers_data}
\begin{tabular}{ll}
\hline
Item & Setting \\
\hline
PDE & One-dimensional viscous Burgers' equation \\
Spatial domain & \(x\in(-3.8,3.8)\) \\
Temporal domain & \(t\in(0,10)\) \\
Initial conditions & \texttt{dual\_exp}, \texttt{exp}, \texttt{sin} \\
High-resolution IQP grid & \(64\times 64\) space-time grid \\
Reference viscosity & \(\nu_0=0.06\) \\
Adaptation viscosities &
\(\{0.11,0.16,0.21,0.26,0.31,0.36,0.41,0.46,0.51\}\) \\
Unseen evaluation viscosities &
\(\{0.13,0.18,0.23,0.28,0.33,0.38,0.43,0.48,0.53\}\) \\
Continuous sample & \((x,t,u)\in\mathbb{R}^3\) \\
Quantization & \(6\) bits per coordinate/value component \\
Bitstring length & \(3\times 6=18\) bits \\
IQP system size & \(18\) qubits \\
Reconstruction & Decode sampled bitstrings back to \((x,t,u)\) values \\
\hline
\end{tabular}
\end{table}

\begin{table}[h]
\centering
\caption{PINN surrogate and physics-informed latent-transfer settings.}
\label{tab:repro_pinn}
\begin{tabular}{ll}
\hline
Item & Setting \\
\hline
Classical surrogate & Latent-conditioned PINN \(u_\Theta(x,t;z)\) \\
PINN latent dimension & \(d_{\mathrm{lat}}=50\) \\
IQP latent dimension & \(d_{\mathrm{lat}}=50\) \\
Main surrogate resolution & \(32\times 32\) \\
Resolution ablations & \(16\times 16\), \(12\times 12\) \\
Transferred latent & PINN latent at \(\nu_0=0.06\) \\
Transferred IQP block & First \(K=50\) IQP parameters \\
Shared-core training & Transferred latent fixed; remaining IQP parameters optimized \\
Latent adaptation & IQP core fixed; first \(K=50\) parameters optimized \\
Adaptation initialization & Previous optimized latent used as warm start \\
\hline
\end{tabular}
\end{table}

\begin{table}[h]
\centering
\caption{IQP circuit and optimization settings.}
\label{tab:repro_iqp}
\begin{tabular}{ll}
\hline
Item & Setting \\
\hline
Number of qubits & \(18\) \\
Gate construction & Pauli-string IQP gates up to locality \(7\) \\
Total IQP parameters & \(63003\) \\
Latent block size & \(K=50\) parameters \\
Shared-core reference viscosity & \(\nu_0=0.06\) \\
Optimizer & Adam \\
Learning rate & \(10^{-4}\) \\
Optimization iterations & \(7500\) \\
Initial random parameter scale & \(10^{-4}\) \\
MMD random operators & \(1500\) \\
MMD samples & \(2000\) \\
Kernel bandwidth & Median heuristic divided by \(5\) \\
Shots for qualitative reconstruction & \(10^6\) \\
\hline
\end{tabular}
\end{table}

\begin{table}[h]
\centering
\caption{Main IQP experiment matrix.}
\label{tab:repro_main_runs}
\begin{tabular}{ll}
\hline
Item & Setting \\
\hline
Initial-condition families & \texttt{dual\_exp}, \texttt{exp}, \texttt{sin} \\
Initialization settings per family &
Random, PINN \(32\times 32\), PINN \(16\times 16\), PINN \(12\times 12\) \\
Total IQP settings & \(3\times 4=12\) \\
Training structure & Core training at \(\nu_0=0.06\), then latent adaptation \\
Adapted training viscosities & \(9\) viscosities \\
Unseen evaluation viscosities & \(9\) viscosities \\
Pairwise latent comparisons & \(\binom{9}{2}=36\) viscosity pairs \\
Reported metrics & MSE, pairwise cosine similarity, Spearman rank correlation \\
\hline
\end{tabular}
\end{table}

\begin{table}[h]
\centering
\caption{Latent-relation ablation settings.}
\label{tab:repro_latent_ablation}
\begin{tabular}{lll}
\hline
Ablation & Quantity Tested & Setting \\
\hline
Viscosity-distance control &
Partial Spearman after controlling for \(D_{ij}=|\nu_i-\nu_j|\) &
\(36\) viscosity pairs per run \\
Viscosity-only baseline &
\(S^{\mathrm{visc}}_{ij}=-|\nu_i-\nu_j|\) &
Compared with \(S^{\mathrm{PINN}}\) and \(S^{\mathrm{IQP}}\) \\
Randomized adaptation order &
Sensitivity to monotonic warm-start order &
\(5\) shuffled seeds per initial condition \\
\hline
\end{tabular}
\end{table}

\begin{table}[h]
\centering
\caption{Randomized adaptation-order ablation settings.}
\label{tab:repro_order_ablation}
\begin{tabular}{ll}
\hline
Item & Setting \\
\hline
Original order &
\(0.11,0.16,0.21,0.26,0.31,0.36,0.41,0.46,0.51\) \\
Random seeds & \(101,102,103,104,105\) \\
Initial conditions & \texttt{dual\_exp}, \texttt{exp}, \texttt{sin} \\
Additional training runs & \(5\times 3=15\) \\
Training procedure & Sequential latent adaptation in shuffled viscosity order \\
Postprocessing & Reorder learned latents to standard viscosity order \\
Comparison & \(\rho(S^{\mathrm{PINN}},S^{\mathrm{IQP}}_{\mathrm{orig}})\) vs.
mean/std of \(\rho(S^{\mathrm{PINN}},S^{\mathrm{IQP}}_{\mathrm{shuf}})\) \\
\hline
\end{tabular}
\end{table}

\begin{table}[h]
\centering
\caption{Conditional DCGAN baseline settings.}
\label{tab:repro_dcgan}
\begin{tabular}{ll}
\hline
Item & Setting \\
\hline
Data representation & Image \(I_\nu=[X,T,U_\nu(X,T)]\) \\
Image shape & \(3\times 64\times 64\) \\
Conditioning variable & Viscosity \(\nu\) \\
Latent dimension & \(50\) \\
Training viscosities &
\(\{0.06,0.11,0.16,0.21,0.26,0.31,0.36,0.41,0.46,0.51\}\) \\
Evaluation viscosities &
\(\{0.13,0.18,0.23,0.28,0.33,0.38,0.43,0.48,0.53\}\) \\
Epochs & \(5000\) \\
Batch size & \(10\) \\
Optimizer & Adam \\
Generator learning rate & \(2\times 10^{-4}\) \\
Discriminator learning rate & \(2\times 10^{-4}\) \\
Adam parameters & \((\beta_1,\beta_2)=(0.5,0.999)\) \\
Real-label smoothing & \(0.1\) \\
Model sizes & Approximately \(37\mathrm{k},64\mathrm{k},96\mathrm{k},278\mathrm{k}\) parameters \\
IQP comparison model & PINN-initialized IQP with \(12\times 12\) surrogate, approximately \(63\mathrm{k}\) parameters \\
Evaluation metric & MSE against numerical Burgers' equation solution \\
\hline
\end{tabular}
\end{table}

\begin{table}[h]
\centering
\caption{Random seeds and numerical reproducibility.}
\label{tab:repro_seeds}
\begin{tabular}{ll}
\hline
Item & Setting \\
\hline
Main IQP runs & Fixed random seeds used for each reported training run \\
Randomized order ablation & Seeds \(101,102,103,104,105\) \\
Classical baselines & Fixed random seed used for each reported model size \\
Stochastic components & Random initialization, MMD estimator, shuffled adaptation order, neural-network training \\
Hardware/backend caveat & Exact bitwise reproducibility is not expected across GPU/software environments \\
Reproducibility level & Protocol-level reproducibility from listed settings, seeds, and evaluation procedure \\
\hline
\end{tabular}
\end{table}

\end{document}